\documentclass[12pt]{article}

\usepackage{cite}

\makeatletter
\if@twoside \oddsidemargin -17pt \evensidemargin 00pt
\else \oddsidemargin 00pt \evensidemargin 00pt
\fi
\expandafter\ifx\csname mathrm\endcsname\relax\def\mathrm#1{{\rm #1}}\fi

\renewcommand{\theequation}{\thesection.\arabic{equation}}
\newcounter{saveeqn}

\@addtoreset{equation}{section}

\makeatother

\def\nl{\nonumber\\}

\def\beq{\begin{equation}}
\def\eeq{\end{equation}}
\def\beqar{\begin{eqnarray}}
\def\eeqar{\end{eqnarray}}
\def\barr#1{\begin{array}{#1}}
\def\earr{\end{array}}
\def\bfi{\begin{figure}}
\def\efi{\end{figure}}
\def\btab{\begin{table}}
\def\etab{\end{table}}
\def\bce{\begin{center}}
\def\ece{\end{center}}
\def\nn{\nonumber}
\def\disp{\displaystyle}
\def\text{\textstyle}

\def\al{\alpha}
\def\be{\beta}
\def\ga{\gamma}
\def\de{\delta}
\def\veps{\varepsilon}

\def\si{\sigma}

\def\De{\Delta}

\def\refeq#1{\mbox{(\ref{#1})}}
\def\reffi#1{\mbox{Fig.~\ref{#1}}}
\def\reffis#1{\mbox{Figs.~\ref{#1}}}
\def\refta#1{\mbox{Table~\ref{#1}}}
\def\reftas#1{\mbox{Tables~\ref{#1}}}
\def\refse#1{\mbox{Sect.~\ref{#1}}}

\def\citere#1{\mbox{Ref.~\cite{#1}}}
\def\citeres#1{\mbox{Refs.~\cite{#1}}}

\newcommand{\eV}{\unskip\,\mathrm{eV}}
\newcommand{\GeV}{\unskip\,\mathrm{GeV}}
\newcommand{\MeV}{\unskip\,\mathrm{MeV}}

\newcommand{\mba}{\unskip\,\mathrm{mb}}

\def\mathswitchr#1{\relax\ifmmode{\mathrm{#1}}\else$\mathrm{#1}$\fi}

\newcommand{\PW}{\mathswitchr W}

\newcommand{\Pe}{\mathswitchr e}

\newcommand{\Pf}{f}

\newcommand{\Pt}{\mathswitchr t}

\newcommand{\Pep}{\mathswitchr {e^+}}
\newcommand{\Pem}{\mathswitchr {e^-}}
\newcommand{\Pepm}{\mathswitchr {e^\pm}}

\def\mathswitch#1{\relax\ifmmode#1\else$#1$\fi}

\newcommand{\MW}{\mathswitch {M_\PW}}

\newcommand{\Me}{\mathswitch {m_\Pe}}

\def\ie{i.e.\ }
\def\eg{e.g.\ }

\renewcommand{\O}{{\cal O}}

\newcommand{\Oa}{\mathswitch{{\cal{O}}(\alpha)}}

\newcommand{\ri}{{\mathrm{i}}}
\newcommand{\re}{{\mathrm{e}}}

\newcommand{\rd}{{\mathrm{d}}}

\newcommand{\M}{{\cal {M}}}

\newcommand{\CM}{{\mathrm{CM}}}

\def\Li{\mathop{\mathrm{Li}_2}\nolimits}

\def\Re{\mathop{\mathrm{Re}}\nolimits}

\def\draftdate{\relax}
\def\mda{\relax}
\def\mua{\relax}
\def\mla{\relax}
\def\draft{
\def\thtystars{******************************}
\def\sixtystars{\thtystars\thtystars}
\typeout{}
\typeout{\sixtystars**}
\typeout{* Draft mode!
         For final version remove \protect\draft\space in source file *}
\typeout{\sixtystars**}
\typeout{}
\def\draftdate{\today}
\def\mua{\marginpar[\boldmath\hfil$\uparrow$]%
                   {\boldmath$\uparrow$\hfil}%
                    \typeout{marginpar: $\uparrow$}\ignorespaces}
\def\mda{\marginpar[\boldmath\hfil$\downarrow$]%
                   {\boldmath$\downarrow$\hfil}%
                    \typeout{marginpar: $\downarrow$}\ignorespaces}
\def\mla{\marginpar[\boldmath\hfil$\rightarrow$]%
                   {\boldmath$\leftarrow $\hfil}%
                    \typeout{marginpar: $\leftrightarrow$}\ignorespaces}
\def\Mua{\marginpar[\boldmath\hfil$\Uparrow$]%
                   {\boldmath$\Uparrow$\hfil}%
                    \typeout{marginpar: $\Uparrow$}\ignorespaces}
\def\Mda{\marginpar[\boldmath\hfil$\Downarrow$]%
                   {\boldmath$\Downarrow$\hfil}%
                    \typeout{marginpar: $\Downarrow$}\ignorespaces}
\def\Mla{\marginpar[\boldmath\hfil$\Rightarrow$]%
                   {\boldmath$\Leftarrow $\hfil}%
                    \typeout{marginpar: $\Leftrightarrow$}\ignorespaces}
\overfullrule 5pt
\oddsidemargin -15mm
\marginparwidth 29mm
}

\makeatletter

\def\eqnarray{\stepcounter{equation}\let\@currentlabel=\theequation
\global\@eqnswtrue
\global\@eqcnt\z@\tabskip\@centering\let\\=\@eqncr
$$\halign to \displaywidth\bgroup\hskip\@centering
  $\displaystyle\tabskip\z@{##}$\@eqnsel&\global\@eqcnt\@ne
  \hskip 2\arraycolsep \hfil${##}$\hfil
  &\global\@eqcnt\tw@ \hskip 2\arraycolsep $\displaystyle\tabskip\z@{##}$\hfil
   \tabskip\@centering&\llap{##}\tabskip\z@\cr}
\def\appendix{\par
 \setcounter{section}{0} \setcounter{subsection}{0}
 \def\thesection{\Alph{section}}}

\makeatother

\newcommand{\gtrless}
{\;\rlap{\raisebox{+.25em}{$>$}}\raisebox{-.25em}{$<$}\;}
\newcommand{\lsim}
{\;\raisebox{-.3em}{$\stackrel{\displaystyle <}{\sim}$}\;}

\newcommand{\beb}{\beta_{\mathrm{b}}}
\newcommand{\gab}{\gamma_{\mathrm{b}}}

\newcommand{\dsidee}{\frac{\rd\si_0}{\rd\bar E'_{\Pe} 
                        \vphantom{\bar{\bar {E'_{\Pe}}}}}}
\newcommand{\dsidea}{\frac{\rd\si_0}{\rd\bar E'_{\ga} 
                        \vphantom{\bar{\bar {E'_{\ga}}}}}}

\begin{document}

\thispagestyle{empty}
\def\thefootnote{\fnsymbol{footnote}}
\setcounter{footnote}{1}
\null
\draftdate\hfill CERN-TH/98-142 \\
\strut\hfill PSI-PR-98-10\\
\strut\hfill hep-ph/9805443
\vskip 0cm
\vfill
\begin{center}
{\Large \boldmath{\bf
Complete ${\cal O}(\alpha)$ QED corrections \\
to polarized Compton scattering}
\par} \vskip 2.5em
{\large
{\sc A.~Denner%
}\\[1ex]
{\normalsize \it Paul Scherrer Institut\\
CH-5232 Villigen PSI, Switzerland}\\[2ex]
{\sc S.~Dittmaier%
}\\[1ex]
{\normalsize \it Theory Division, CERN\\
CH-1211 Geneva 23, Switzerland}\\[2ex]
}
\par \vskip 1em
\end{center}\par
\vskip .0cm \vfill {\bf Abstract:} \par 
The complete QED corrections
of ${\cal O}(\alpha)$ to polarized Compton scattering are calculated
for finite electron mass and including the real corrections induced by
the processes $\Pem\gamma\to\Pem\ga\ga$ and
$\Pem\gamma\to\Pem\Pem\Pep$. All relevant formulas are listed in a
form that is well suited for a direct implementation in computer
codes. We present a detailed numerical discussion of the 
${\cal O}(\alpha)$-corrected cross section and the 
left--right asymmetry in
the energy range of present and future Compton polarimeters, which are
used to determine the beam polarization of high-energetic
$\Pe^\pm$ beams. For
photons with energies of a few eV and electrons with SLC energies or smaller, 
the corrections are of the order of a few per mille.
In the energy range of future $\Pep\Pem$ colliders, however, they reach
1--2\% and cannot be neglected in a precision polarization measurement.
\par
\vskip 1cm
\noindent
CERN-TH/98-142 \\
May 1998
\par
\null
\setcounter{page}{0}
\clearpage
\def\thefootnote{\arabic{footnote}}
\setcounter{footnote}{0}

\section{Introduction}

The sensitivity of Compton scattering, $\Pepm\gamma\to\Pepm\gamma$, to the
polarization of both the electron and the photon in the initial state
renders this process 
well suited for the determination of the
polarization of electron (or positron) beams in high-energy collider
experiments. Usually, in a Compton polarimeter,
circularly polarized laser
light is brought to collision with the high-energetic $\Pe^\pm$ beam,
and the degree of beam polarization is deduced from the left--right
asymmetry with respect to the polarization of the
laser photons. The high precision of such
polarization measurements, which is typically 
of the order of 1\% \cite{sld,cebaf,tjnaf,tesla}, 
sets the necessary level of accuracy in theoretical predictions to a few per
mille. Therefore, it is indispensable to control radiative corrections.

In the low-energy limit, the QED corrections to 
the Compton scattering cross section are known to vanish 
to all orders of perturbation theory \cite{th50,di97}, 
i.e.\ the relative corrections are suppressed by a factor of the
electron velocity $\beta$ for small $\beta$.
This fact, which is known as 
Thirring's theorem, is due to the definition
of the electromagnetic charge in the Thomson limit and electromagnetic
gauge invariance. 
On the other hand, the relative corrections 
to the polarization asymmetry are not suppressed with $\beta$, since the
asymmetry itself vanishes for $\beta\to 0$.
Therefore, the QED corrections to the asymmetry are expected to be 
of the order of $\alpha/\pi$ times some numerical factor, i.e.\ of 0.1--1\%,
even for small scattering energies.

The first calculation of the
virtual and real soft-photonic QED corrections in ${\cal O}(\alpha)$ was
performed by Brown and Feynman \cite{br52} for the unpolarized cross
section in 1952. In the same year, the
amplitudes for the corresponding hard-photonic
bremsstrahlung, so-called double Compton scattering,
$\Pem\gamma\to\Pem\gamma\gamma$, were given by Mandl and Skyrme
\cite{ma52}. The investigation of radiative
corrections to polarized particles started much later: 
in 1972 the one-loop and real soft-photonic corrections were presented by 
Milton, Tsai, and De~Raad \cite{mi72}. The helicity amplitudes for
hard-photon radiation were given by G\'ongora-T.
and Stuart \cite{go89} in 1989.
In the same year the first numerical evaluation of polarized Compton
scattering was published by H.~Veltman \cite{ve89}, who, in particular,
studied the impact of QED corrections on the beam-polarization
measurement at SLAC. 
A recent evaluation by Swartz \cite{sw97}, which is
based on the (corrected) formulas of \citeres{mi72, go89}, could not
confirm Veltman's results. 
Even though both authors find corrections of
the order of $0.1$--0.2\% for the SLAC polarimeter, which is small with
respect to the experimental uncertainty of about $0.7\%$
\cite{sld}, it is necessary to settle this issue. 
{}For a Next Linear Collider (NLC) with $\Pe^\pm$ beams of $500\GeV$, 
Swartz finds QED corrections of 1--2\%, definitely requiring
their inclusion in the determination of the beam polarization.

In view of this situation, 
we have performed a completely
independent calculation of the ${\cal O}(\alpha)$ QED corrections to
polarized Compton scattering, the analytical and numerical results of
which are presented in this paper. 
In order to facilitate the use of our formulas, the presentation
is held at a rather detailed level. As done in \citere{sw97}, we also
include the real corrections induced by the process
$\Pem\gamma\to\Pem\Pem\Pep$, which will be relevant at beam energies of
future colliders if not the photons but the electrons of the final state
are detected. 

Because Compton scattering is a pure QED process in lowest order,
the electroweak radiative corrections can be split into a QED part and a
genuinely weak part in a gauge-invariant way, just by selecting the QED
diagrams and the remaining ones at each loop level. Since Thirring's
low-energy theorem is also valid in the electroweak Standard Model
\cite{di97}, also the weak corrections vanish in the low-energy limit.
At any centre-of-mass (CM) energy 
$E_{\mathrm{CM}}$ that is far below the electroweak 
bosons' masses $\MW$, etc., the weak corrections are 
suppressed by factors of $E_{\mathrm{CM}}^2/\MW^2$. 
This feature can be observed directly
by inspecting the weak one-loop corrections 
that were explicitly
calculated in \citeres{de93,di94}. Owing to this suppression, the weak
corrections are irrelevant for a Compton polarimeter, where
$E_{\mathrm{CM}} = {\cal O}(\Me)$ and thus $E_{\mathrm{CM}}^2\ll\MW^2$.
However, for CM
energies above the weak scale $\MW$, i.e.\ in the range of future
$\Pe^\pm\gamma$ colliders, these corrections become large, reaching
5--10\%\ in the TeV range.

This article is organized as follows:
in \refse{se:locs} we set some conventions and present the necessary
formulas for the lowest-order predictions. Section~\ref{se:virt} contains
our analytical results for the virtual one-loop corrections. The
correction factor for soft-photonic bremsstrahlung as well as the
helicity amplitudes for 
$\Pem\gamma\to\Pem\gamma\gamma,\Pem\Pem\Pep$ are
given in \refse{se:real}. In \refse{se:numres} we discuss our numerical
results, which include studies for Compton polarimeters for
$\Pe^\pm$-beam energies of 4--8$\GeV$ (CEBAF), $50\GeV$ (SLAC), and
$500\GeV$ (NLC).

\section{Conventions and lowest-order cross-sections}
\label{se:locs}

The momenta and polarizations for the Compton process are assigned as
follows:
\beq
\Pem(p,\sigma) + \gamma(k,\lambda) \; \longrightarrow \;
\Pem(p',\sigma') + \gamma(k',\lambda'),
\eeq
where the helicities take the values $\sigma,\sigma'=\pm 1/2$ and
$\lambda,\lambda'=\pm 1$. For brevity the helicities are often
indicated by their sign.
We have chosen incoming electrons for definiteness; the results for
incoming positrons follow by charge conjugation.

\begin{figure}
\centerline{
\setlength{\unitlength}{1cm}
\begin{picture}(10.5,2.5)
\put(-2.5,-14.5){\includegraphics{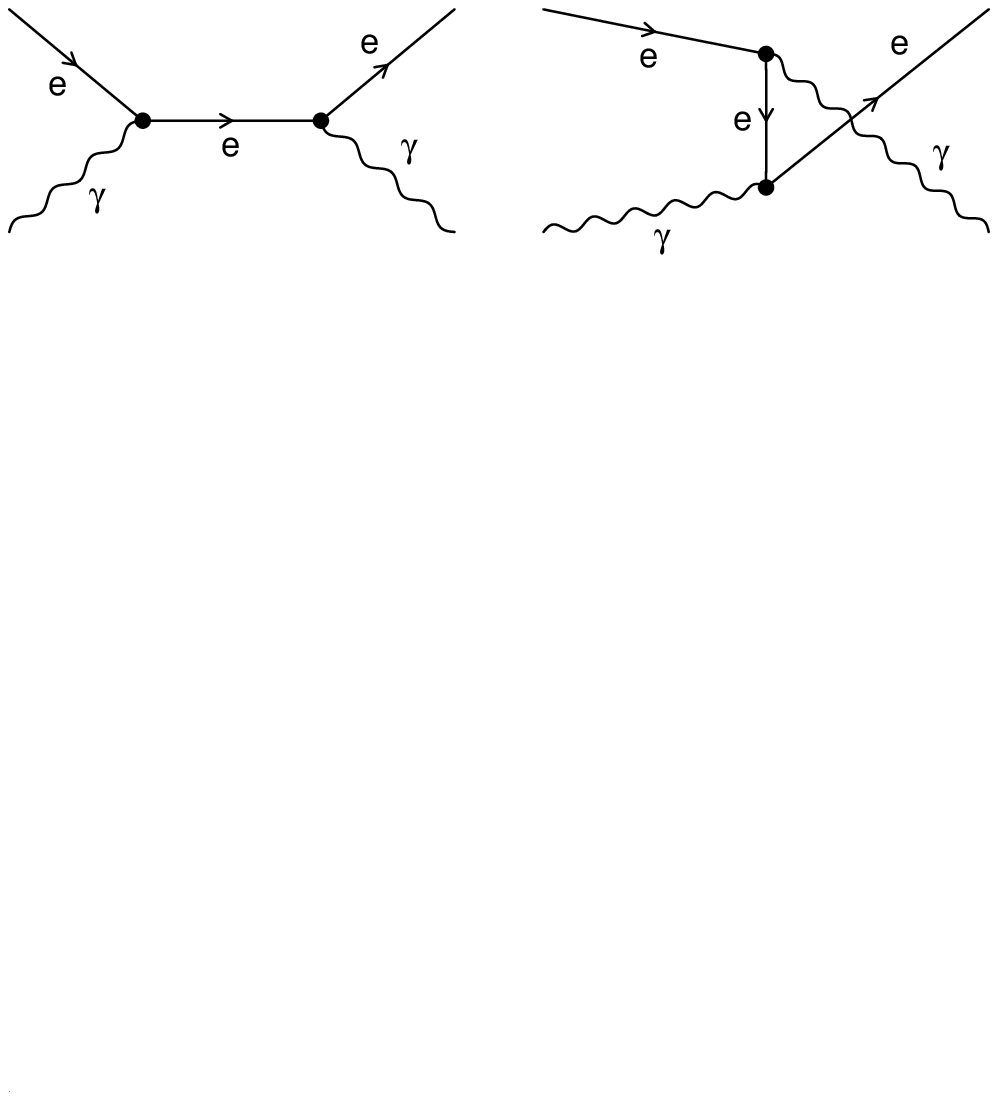}}
\end{picture} }
\caption{Tree diagrams for $\Pem\gamma\to\Pem\gamma$}
\label{fi:eaea0diags}
\end{figure}

We first specify the kinematics in the CM system, where all momenta are
fixed by the energy of one particle
and the scattering angle in the scattering
plane. Denoting the photon energy by $E_\gamma$, the electron energy
$E_\Pe$ and the electron velocity $\beta$ are given by
\beq\label{Eebeta}
E_\Pe=\sqrt{E_\gamma^2+\Me^2}, \qquad
\beta=E_\gamma/E_\Pe.
\label{eq:Ee}
\eeq
Identifying the beam axes with the $z$ direction and taking the 
$x$--$z$ plane as scattering plane, the momenta read
\beqar
p^\mu         &=& \parbox{5cm}{$E_\Pe(1,0,0,\beta),$}
k^\mu          =  E_\gamma(1,0,0,-1),  \nn\\
p^{\prime\mu} &=& \parbox{5cm}{$E_\Pe(1,\beta\sin\theta,0,\beta\cos\theta),$}
k^{\prime\mu}  =  E_\gamma(1,-\sin\theta,0,-\cos\theta).
\label{eq:momCMS}
\eeqar
The Mandelstam variables are defined as usual:
\begin{eqnarray}\label{Mandelstam}
s &=& (p+k)^2 \;=\; (p'+k')^2 \;=\; E_{\CM}^2 = (E_\Pe+E_\gamma)^2,
\nn\\
t &=& (p-p')^2 \;=\; (k-k')^2 \;=\; \text
-4E_\gamma^2\sin^2\big(\frac{\theta}{2}\big), \nn\\
u &=& (p-k')^2 \;=\; (p'-k)^2 \;=\; \text
(E_\Pe-E_\gamma)^2-4E_\gamma^2\cos^2\big(\frac{\theta}{2}\big).
\label{stu}
\end{eqnarray}
Moreover, it is convenient to introduce the shorthands
\beqar
s_m &=& \parbox{4cm}{$s-\Me^2 = 2E_\gamma\sqrt{s},$} 
u_m = u-\Me^2 = 
-2E_\gamma\biggl[2E_\gamma\cos^2\big(\text\frac{\theta}{2}\big)
+\frac{\Me^2}{\sqrt{s}}\biggr], 
\nn\\
t_m &=& \parbox{4cm}{$t-4\Me^2,$} 
r = \sqrt{\Me^4-su} = s_m \cos\big(\text\frac{\theta}{2}\big).
\hspace{2em}
\eeqar
{}For the photon polarization vectors of helicity eigenstates we choose
\beqar
\veps^\mu(k,\lambda=\pm 1) &=& \frac{1}{\sqrt{2}}(0,1,\mp \ri,0),
\nn\\
\veps^{\prime*\mu}(k',\lambda'=\pm 1) &=& 
\frac{1}{\sqrt{2}}(0,\cos\theta,\pm \ri,-\sin\theta).
\label{eq:poldef}
\eeqar
{}For the helicity eigenspinors of the electrons we refer to
\citere{di98}. The discrete symmetries of QED,
actually parity and the combination of charge conjugation and time
reversal, reduce the number of independent helicity amplitudes for
$\Pem\gamma\to\Pem\gamma$ to six. In lowest order these amplitudes,
$\M_0(\sigma,\lambda,\sigma',\lambda')$, receive contributions from the
two tree diagrams shown in \reffi{fi:eaea0diags} and are given by
\beqar
\M_0({\pm}{\pm}{\pm}{\pm}) &=& 
8\pi\alpha\frac{r(s_m^2+\Me^2 t)}{s_m^2 u_m},
\nn\\[.5em]
\M_0({\pm}{\mp}{\pm}{\mp}) &=& 
8\pi\alpha\frac{r^3}{s_m^2 u_m},
\nn\\[.5em]
\M_0({\pm}{\mp}{\mp}{\pm}) &=& 
\pm 8\pi\alpha\frac{\Me st\sqrt{{-}t}}{s_m^2 u_m},
\nn\\[.5em]
\M_0({\pm}{\mp}{\mp}{\mp}) &=& \M_0({\pm}{\pm}{\mp}{\pm}) = 
\pm 8\pi\alpha\frac{\Me r^2\sqrt{-t}}{s_m^2 u_m},
\nn\\[.5em]
\M_0({\pm}{\pm}{\pm}{\mp}) &=& \M_0({\pm}{\mp}{\pm}{\pm}) = 
8\pi\alpha\frac{\Me^2 rt}{s_m^2 u_m},
\nn\\[.5em]
\M_0({\pm}{\pm}{\mp}{\mp}) &=& 
\pm 8\pi\alpha\frac{\Me^3 t\sqrt{-t}}{s_m^2 u_m},
\label{eq:amplo}
\eeqar
where $\alpha$ is the fine-structure constant.

Usually, for $\Pe^\pm\gamma$ collisions 
the CM system and the laboratory
(LAB) frame do not coincide. Throughout this paper we assume that 
the CM and LAB frames are related by a boost along the beam axes. 
This boost is completely fixed by specifying the energies $\bar E_\Pe$ and 
$\bar E_\gamma$ of the incoming electron and photon in the LAB frame.
Actually, in practice the angle between the incoming beams in the LAB system
deviates from $180^\circ$ by a very small angle 
$\alpha_c$ of some mrad, but the impact of this effect is
negligible for present and future polarimeters. 
More details about the situation with non-zero $\alpha_c$ are given in
the appendix.

In terms of LAB
variables, which are marked by bars, the particle momenta read
\beqar
p^\mu         &=& \parbox{5.5cm}{$\bar E_\Pe(1,0,0,\bar\beta),$}
k^\mu          =  \bar E_\gamma(1,0,0,-1),  \nn\\
p^{\prime\mu} &=& \parbox{5.5cm}{$\bar E'_\Pe(1,
\bar\beta'\sin\bar\theta'_\Pe,0,\bar\beta'\cos\bar\theta'_\Pe),$}
k^{\prime\mu}  =  \bar E'_\gamma(1,\sin\bar\theta'_\gamma,0,
\cos\bar\theta'_\gamma),
\label{eq:momLAB}
\eeqar
where
\beq
\bar\beta  = \sqrt{1-\Me^2/\bar E_\Pe^2}, \qquad
\bar\beta' = \sqrt{1-\Me^2/\bar E^{\prime 2}_\Pe}
\eeq
are the velocities of the respective electrons. The relation between 
quantities in the different frames are most conveniently written in
terms of the boost parameters
\beq
\beb = \frac{\bar\beta\bar E_\Pe-\bar E_\gamma}{\bar E_\Pe+\bar
  E_\gamma}, \qquad
\gab = \frac{1}{\sqrt{1-\beb^2}} = \frac{\bar E_\Pe+\bar E_\gamma}{\sqrt{s}},
\eeq
where
\beq
s = \Me^2+2\bar E_\gamma\bar E_\Pe(1+\bar\beta).
\eeq
The photon energy in the CM system is given by
\beq
E_\gamma = \bar E_\gamma \gab(1+\beb),
\eeq
from which $E_\Pe$ and $\beta$ follow 
through \refeq{eq:Ee}. The scattering
angle $\theta$ in the CM system is related to the energy $\bar E'_\Pe$
of the outgoing electron and to the energy $\bar E'_\gamma$ of the
outgoing photon in the LAB frame by
\beq
\cos\theta = \frac{\bar E'_\Pe-\gab E_\Pe}{E_\gamma \gab \beb}
= \frac{\gab E_\gamma-\bar E'_\gamma}{E_\gamma \gab \beb}.
\label{eq:thboost}
\eeq
The extreme values of $\bar E'_\Pe$ and $\bar E'_\gamma$ correspond to
the scattering into forward and backward directions:
\beqar
\bar E'_{\Pe,\max} &=& \bar E_\Pe+\bar E_\gamma-\bar E'_{\gamma,\min}
= \bar E_\Pe, \nn\\
\bar E'_{\gamma,\max} &=& \bar E_\Pe+\bar E_\gamma-\bar E'_{\Pe,\min}
= \frac{E_\gamma^2}{\bar E_\gamma},
\eeqar
provided that $\beb<\beta$. This condition, which is obviously fulfilled 
for Compton polarimeters, ensures that the scattered electron cannot be
at rest in the LAB frame.

Note that the amplitudes \refeq{eq:amplo} refer to helicity eigenstates
in the CM frame, and that the property of being a helicity eigenstate is
frame-dependent for massive particles. In our case, however, the
incoming electron is in a state of definite helicity both in the CM and
in the LAB system, because the boost that relates 
the two frames goes along the direction of flight for this particle.
The outgoing electron is in general not
in a helicity eigenstate in the LAB frame. In practice this does not
matter, as the spins of the outgoing particles are not measured.

The basic observable is
the differential lowest-order cross section in the CM system,
\beq
\frac{\rd\sigma_{0}(P_\Pe,P_\gamma)}{\rd\cos\theta} = 
\frac{1}{32\pi s} \sum_{\sigma,\lambda,\sigma',\lambda'}
\frac{1}{4}(1+2\sigma P_\Pe)(1+\lambda P_\gamma)
\, |\M_0(\sigma,\lambda,\sigma',\lambda')|^2, 
\label{eq:bdsdth}
\eeq
where $P_\Pe$ and $P_\gamma$ denote the respective degrees of beam
polarization, which do not change by the boost to the LAB frame.
Using \refeq{eq:thboost} this leads us, for instance, directly to the
electron-energy distribution in the LAB frame:
\beq
\frac{\rd\sigma_{0}(P_\Pe,P_\gamma)}{\rd\bar E'_\Pe} =
\frac{1}{E_\gamma \gab\beb} 
\left. \frac{\rd\sigma_{0}(P_\Pe,P_\gamma)}{\rd\cos\theta}
\right|_{\cos\theta = \frac{\bar E'_\Pe-\gab E_\Pe}{E_\gamma \gab\beb}}.
\label{eq:beedist}
\eeq
The most important quantity for a Compton polarimeter is the
polarization asymmetry $A_{\mathrm{LR}}(P_\Pe)$ with respect to
incoming laser photons of opposite polarization:
\beq
A_{\mathrm{LR}}(P_\Pe) = 
\frac{\Delta\sigma(P_\Pe,-1)-\Delta\sigma(P_\Pe,+1)}
{\Delta\sigma(P_\Pe,-1)+\Delta\sigma(P_\Pe,+1)},
\eeq
where the defining cross section $\Delta\sigma(P_\Pe,P_\gamma)$ can be
the integrated cross section or any distribution measured by the
detector. As long as effects of the weak interaction can be neglected
(see introduction), parity is an exact symmetry, and we have
$\Delta\sigma(P_\Pe,P_\gamma)=\Delta\sigma(-P_\Pe,-P_\gamma)$. In this
case we get
\beq\label{asymm}
A_{\mathrm{LR}}(P_\Pe) = P_\Pe A_{\mathrm{LR}}(+1),
\eeq
and $P_\Pe$ can be determined by the measured value 
of $A_{\mathrm{LR}}(P_\Pe)$ divided by the theoretical prediction for 
$A_{\mathrm{LR}} \equiv A_{\mathrm{LR}}(+1)$.

Note that \refeq{eq:bdsdth} makes use of rotational invariance with
respect to the beam axes, which is also assumed in the above treatment
of the kinematics. This procedure is adequate for helicity states, but
should be generalized if one is interested in transverse polarizations,
which are not considered in this paper. It is 
quite easy to deduce the amplitudes for any polarization
configuration from our helicity amplitudes, both for the lowest order 
and for the corrections described in the next sections.

\section{Virtual corrections}
\label{se:virt}

{}For the virtual corrections the kinematics is the same as in lowest order.
Lowest-order amplitudes and loop amplitudes 
simply add up to the full amplitudes $\M$. The squared amplitudes $|\M|^2$ 
are expanded in $\alpha$. 
In ${\cal O}(\alpha)$ precision this leads to
$|\M|^2=|\M_0|^2+2\Re\{\M_0\M_1^*\}+{\cal O}(\alpha^2)$, 
where $\M_1$ is the sum of all one-loop 
diagrams. Therefore, in one-loop approximation the virtual correction to
the differential Born cross section \refeq{eq:bdsdth} in the CM system reads
\beq
\frac{\rd\sigma_{\mathrm{V}}(P_\Pe,P_\gamma)}{\rd\cos\theta} = 
\frac{1}{32\pi s} \sum_{\sigma,\lambda,\sigma',\lambda'}
\frac{1}{4}(1+2\sigma P_\Pe)(1+\lambda P_\gamma)
\, 2\Re\{ \M_0(\sigma,\lambda,\sigma',\lambda')
\M_1^*(\sigma,\lambda,\sigma',\lambda') \}. 
\label{eq:vdsdth}
\eeq
The transition to other distributions like the one of \refeq{eq:beedist}
is obtained as in lowest order.

\begin{figure}
\centerline{
\setlength{\unitlength}{1cm}
\begin{picture}(16,5.5)
\put(-2.5,-12.0){\includegraphics{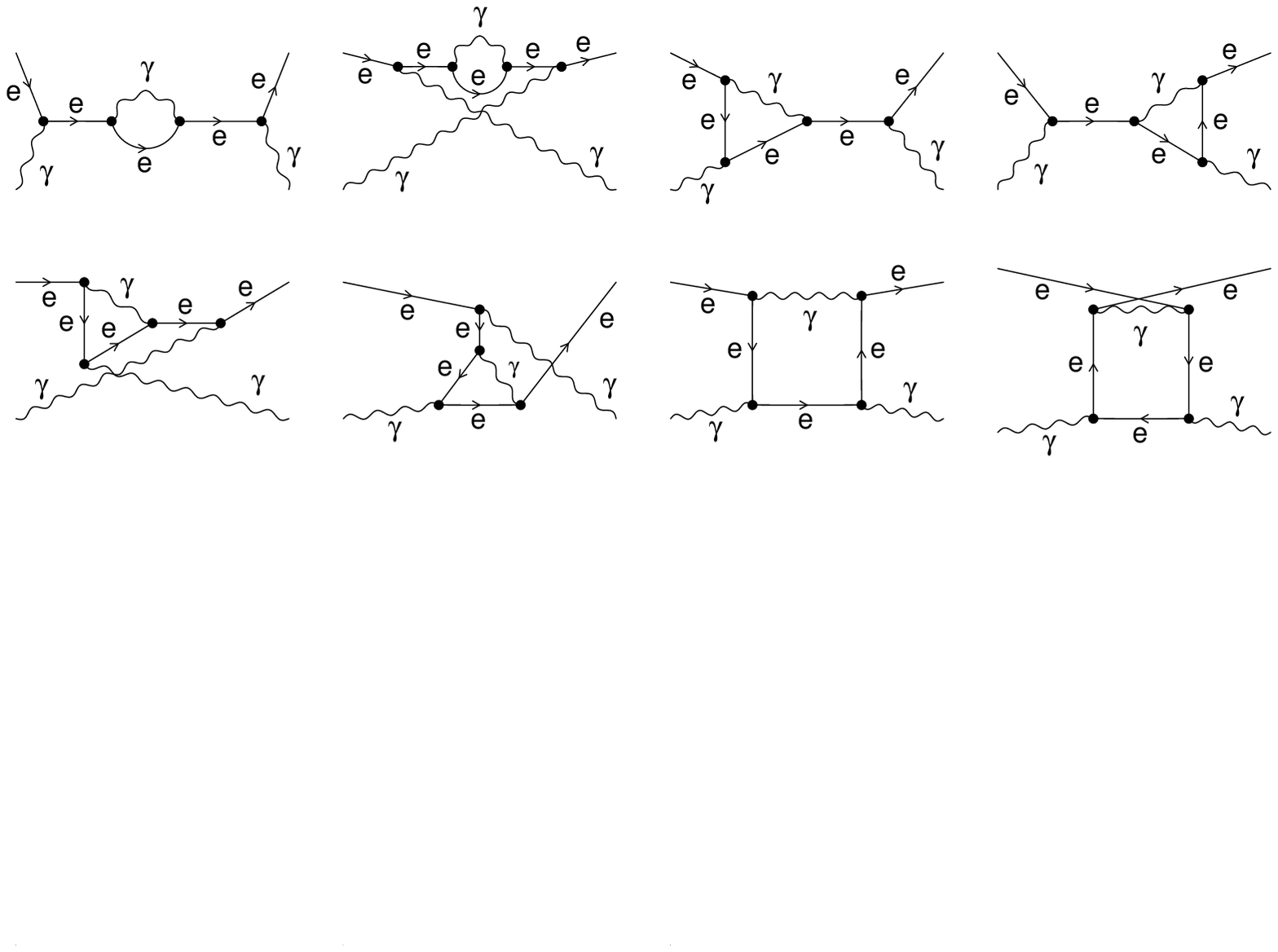}}
\end{picture} }
\caption{One-loop QED diagrams for $\Pem\gamma\to\Pem\gamma$}
\label{fi:eaea1diags}
\end{figure}
The one-loop diagrams, which are depicted in \reffi{fi:eaea1diags},
have been calculated using the standard technique
described in \citere{ADHab}. This means that one-loop tensor integrals
are algebraically reduced to scalar integrals 
\`a la Passarino and Veltman
\cite{pa79}, and the scalar integrals are analytically integrated using
the general methods described in \citere{th79}. The algebraic part of
the calculation has been performed in two different ways. The first
calculation is based on the computer algebra packages {\sl FeynArts}
\cite{FA} and {\sl FeynCalc} 
\cite{FC}; the former generates the one-loop amplitudes,
the latter performs several kinds of algebraic manipulations
(contractions, application of Dirac's equation, etc.) and reduces the
tensor integrals to scalar integrals. The second calculation makes
direct use of the existing one-loop results for the process
$\gamma\gamma\to\Pt\bar\Pt$ \cite{de95}, which were derived once with
and once without {\sl FeynCalc}. 
The reaction $\gamma\gamma\to\Pt\bar\Pt$ is
related to Compton scattering via crossing symmetry, i.e.\ one simply
has to reverse the momenta of the anti-top quark and of one of the photons,
to replace the corresponding spinor and polarization vector, and to
substitute the top-quark parameters (charge, mass, colour) by the
corresponding parameters of the electron. The final step of evaluating the
spinor chains for definite helicity states has also been carried out in
different ways.

In QED the one-loop amplitudes $\M_1(\sigma,\lambda,\sigma',\lambda')$ 
are related by the same discrete symmetries as the Born amplitudes
\refeq{eq:amplo}. In terms of scalar integrals the independent $\M_1$
are given by
\beqar
\lefteqn{
\M_1({+}{+}{+}{+}) = 2\alpha^2\frac{r}{s_m}\biggl\{
      \frac{s_m^2u+\Me^2t^2-2\Me^4t}{s u_m u}
    + 4\Me^2\frac{s_m^2+\Me^2 t}{s_m u_m} B'
    + \frac{2t}{t_m} B_{mm}(t) } &&
\nn\\ && {}
    + \Me^2\frac{5s^2-2\Me^2 s+\Me^4}{s s_m^2} B_{0m}(s)
    + \frac{r^2u+2tu^2+4\Me^2u^2+\Me^4t}{u u_m^2} B_{0m}(u)
\nn\\ && {}
    + \frac{1}{r^2}\biggl[ \,
        - t(s^2+t^2+\Me^2t-9\Me^4) C_{mmm}(t)
        + t_m(s_m^2-t^2-\Me^2t) C_{m0m}(t)
\hspace{3em}
\nn\\ && {} \phantom{{}+\frac{1}{r^2}\biggl[}
        - 2s_m[s^2-\Me^2(4s+u)] C_{0mm}(s)
        - 2u_m(t^2+3\Me^2s-7\Me^4) C_{0mm}(u)
\nn\\ && {} \phantom{{}+\frac{1}{r^2}\biggl[}
        + [s_m^3t-\Me^2s_m(2s_m^2+s_m t_m-t t_m)+2\Me^4t^2-4\Me^6t] 
          D_{m0mm}(t,s)
\nn\\ && {} \phantom{{}+\frac{1}{r^2}\biggl[}
        + [t^3u_m+\Me^2(2t^3+s_m^2u_m-s_m u_m^2)-2\Me^4(s_m^2-3u_m^2)-4\Me^6t]
          D_{m0mm}(t,u)
             \, \biggr] \biggr\},
\nn\\[.5em]
\lefteqn{
\M_1({+}{-}{+}{-}) = 2\alpha^2\frac{r}{s_m}\biggl\{
    - \frac{u_m}{u} + 4\Me^2\frac{r^2}{s_m u_m} B' 
    + 2t\frac{s_m^2+\Me^2(3s_m-u_m)}{r^2 t_m} B_{mm}(t) }
\nn\\ && {}
    - (s+\Me^2)\frac{s_m^2(2s_m+3u)+5\Me^2s_m u_m-2\Me^4t}{r^2s_m^2} B_{0m}(s)
\nn\\ && {}
    - \Me^2\frac{3s_m u u_m^2-2u_m^4-\Me^2(4r^2u+u_m^3)}{r^2 u u_m^2} B_{0m}(u)
\nn\\ && {}
    + \frac{1}{r^4}\biggl[ \,
        - t[2r^2st_m+s^2s_m^2+\Me^2(s_m^3-ss_m u_m+s_m uu_m-uu_m^2)] C_{mmm}(t)
\nn\\ && {} \phantom{{}+\frac{1}{r^4}\biggl[}
        + t_m[s_m^3(s_m+2t)-\Me^2t^2(4s_m+t)-\Me^4t^2] C_{m0m}(t)
\nn\\ && {} \phantom{{}+\frac{1}{r^4}\biggl[}
        - 2s_m[s^2t t_m-\Me^2s s_m(3s_m+2t)+\Me^4s_m u_m] C_{0mm}(s)
\nn\\ && {} \phantom{{}+\frac{1}{r^4}\biggl[}
        + 2u_m[r^2su+\Me^2u(s_m t+u_m^2)+\Me^4(2s_m^2+3st)] C_{0mm}(u)
\nn\\ && {} \phantom{{}+\frac{1}{r^4}\biggl[}
        + (s+\Me^2)[ s^2t^3+\Me^2s_m^2(2s_m^2+3s_m t-2t^2)-5\Me^4s_m t^2
          -2\Me^6t^2] D_{m0mm}(t,s)
\nn\\ && {} \phantom{{}+\frac{1}{r^4}\biggl[}
        + [s_m^2t u_m^3+\Me^2u_m(3t^3u+t^2u_m^2-5tu_m^3-2u_m^4)
         -\Me^4u_m^2(5t^2+6tu_m-4u_m^2)
\nn\\ && {} \phantom{{}+\frac{1}{r^4}\biggl[}
         +2\Me^6t^2(t - 2u)] D_{m0mm}(t,u)
                      \, \biggr] \biggr\}, 
\nn\\[.5em]
\lefteqn{
\M_1({+}{-}{-}{+}) = 4\alpha^2\frac{\Me\sqrt{-t}}{s_m}\biggl\{ \,
      \frac{s_m}{u_m} + 2\Me^2\frac{st}{s_mu_m} B'      
    + s\frac{3s-\Me^2}{s_m^2} B_{0m}(s)                 }
\nn\\ && {}
    + \frac{3su+\Me^2s-2\Me^4}{u_m^2} B_{0m}(u) 
    - 2\Me^2 C_{mmm}(t) - 2\Me^2\frac{t}{u_m} C_{0mm}(u)
\nn\\ && {}
    - \Me^2t D_{m0mm}(t,u) + (st-3\Me^2s+\Me^4)[D_{m0mm}(t,s)+D_{m0mm}(t,u)]
                     \, \biggl\},
\nn\\[.5em]
\lefteqn{
\M_1({+}{-}{-}{-}) = 2\alpha^2\frac{\Me\sqrt{-t}}{s_m}\biggl\{
    - \frac{r^2}{u u_m} + 4\Me^2\frac{r^2}{s_m u_m} B'  
    + 2\frac{s-u}{t_m} B_{mm}(t) 
    + 4\frac{s^2}{s_m^2} B_{0m}(s)              }
\nn\\ && {}
    - \frac{r^2u_m+2u u_m^2-4\Me^4u}{u u_m^2} B_{0m}(u)
\nn\\ && {}
    + \frac{1}{r^2}\biggl[ \,
      - t(2st+u u_m-7\Me^2s-\Me^4) C_{mmm}(t)
      + t_m(2s_m^2-u_m^2-\Me^2t) C_{m0m}(t)
\nn\\ && {} \phantom{{}+\frac{1}{r^2}\biggl[}
      - 2s_m(r^2-2s^2-2\Me^2s) C_{0mm}(s)
      + 2u_m(2s+t)(u+\Me^2) C_{0mm}(u)
\nn\\ && {} \phantom{{}+\frac{1}{r^2}\biggl[}
      - (s_m^3t-s^2t t_m-4\Me^2s s_m^2-\Me^2st^2) D_{m0mm}(t,s)
\nn\\ && {} \phantom{{}+\frac{1}{r^2}\biggl[}
      + (t^2u^2+2t_m u_m^3+4\Me^2u u_m^2+\Me^4t t_m) D_{m0mm}(t,u)
                       \, \biggr] \biggr\},
\nn\\[.5em]
\lefteqn{
\M_1({+}{+}{+}{-}) = 2\alpha^2\frac{\Me^2r}{s_m u_m}\biggl\{
      \frac{s-u}{\Me^2} + 4\Me^2\frac{t}{s_m} B'
    + 4\frac{s u_m}{s_m^2} B_{0m}(s) 
    + 2\frac{(2u-t)}{u_m} B_{0m}(u)     }
\nn\\ && {}
    + \frac{1}{r^2}\biggl[ \,
    - 2u_m[s_m(s-u)+2\Me^2t] C_{mmm}(t)
    - 2st u_m C_{0mm}(s)
    - 2t(r^2-\Me^2u_m) C_{0mm}(u)
\nn\\ && {} \phantom{{}+\frac{1}{r^2}\biggl[}
    - u_m[st(s+2u)+4\Me^2r^2-3\Me^4t] [D_{m0mm}(t,s)+D_{m0mm}(t,u)]
\nn\\ && {} \phantom{{}+\frac{1}{r^2}\biggl[}
    - t^2u_m(s+\Me^2) D_{m0mm}(t,u)
               \, \biggr] \biggr\},
\nn\\[.5em]
\lefteqn{
\M_1({+}{+}{-}{-}) = 4\alpha^2\frac{\Me^3\sqrt{-t}}{s_m^2}\biggl\{ \,
    - \frac{s_m}{\Me^2} + 2\Me^2\frac{t}{u_m} B'
    + \frac{s+\Me^2}{s_m} B_{0m}(s) + s_m\frac{3u-\Me^2}{u_m^2} B_{0m}(u) }
\nn\\ && {}
    - 2s_m[ C_{mmm}(t)+C_{0mm}(s)-C_{0mm}(u)] - s_m(2s+u_m)D_{m0mm}(t,s) 
\nn\\ && {}
    - s_m(s+\Me^2) D_{m0mm}(t,u)        
                \, \biggr\}.
\label{eq:amponel}
\eeqar
The scalar integrals read explicitly
\beqar
B' &=& \frac{\partial B_0}{\partial p^2}(p^2,\Me,\lambda)\bigg|_{p^2=\Me^2} =
-\frac{1}{\Me^2}\biggl[1+\ln\biggl(\frac{\lambda}{\Me}\biggr)\biggr],
\nn\\[.5em]
B_{mm}(t) &=& B_0(t,\Me,\Me) - B_0(0,\Me,\Me) - 2 = \beta_t\ln(x_t),
\nn\\[.5em]
B_{0m}(v) &=& B_0(v,0,\Me) - B_0(0,0,\Me) - 1 =
-\frac{v_m}{v}\ln\biggl(-\frac{v_m+\ri\epsilon}{\Me^2}\biggr),
\nn\\[.5em]
C_{mmm}(t) &=& C_0(0,0,t,\Me,\Me,\Me) = \frac{1}{2t}\ln^2(x_t),
\nn\\[.5em]
C_{m0m}(t) &=& C_0(\Me^2,\Me^2,t,\Me,\lambda,\Me) 
\nn\\ 
&=& \frac{1}{\beta_t t}\biggl[ \ln\biggl(\frac{\lambda^2}{\Me^2}\biggr)\ln(x_t)
+\frac{1}{2}\ln^2(x_t)+2\Li(1+x_t)-\frac{\pi^2}{2}-2\pi\ri\ln(1+x_t) \biggr],
\nn\\[.5em]
C_{0mm}(v) &=& C_0(\Me^2,0,v,0,\Me,\Me) 
\nn\\ &=& 
\frac{1}{v_m}\biggl[ 
\Li\biggl(-\frac{v_m+\ri\epsilon}{\Me^2}\biggr) 
+\ln\biggl(\frac{v+\ri\epsilon}{\Me^2}\biggr)
\ln\biggl(-\frac{v_m+\ri\epsilon}{\Me^2}\biggr)
\biggr],
\nn\\[.5em]
D_{m0mm}(t,v) &=& D_0(\Me^2,\Me^2,0,0,t,v,\Me,\lambda,\Me,\Me)
\nn\\
&=& \frac{1}{\beta_t tv_m}\biggl[
2\ln(x_t)\ln\biggl(\frac{\lambda\Me}{-v_m-\ri\epsilon}\biggr)-2\Li(1-x_t)
\nn\\ && \phantom{\frac{1}{\beta_t t(\Me^2-v)}\biggl[} {}
+2\Li(1+x_t)-\frac{\pi^2}{2}-2\pi\ri\ln(1+x_t) \biggr],
\label{eq:scalint}
\eeqar
with the dilogarithm $\Li(x)=-\int_0^1\rd t\ln(1-xt)/t$ and
\beq
v_m = v-\Me^2, \qquad
\beta_t = \sqrt{1-\frac{4\Me^2}{t+\ri\epsilon}}, \qquad
x_t = \frac{\beta_t-1}{\beta_t+1}.
\eeq
In \refeq{eq:scalint}, $\lambda$ denotes an infinitesimal photon mass,
which is chosen as IR regulator and 
drops out after adding 
soft-bremsstrahlung corrections. The quantity $\ri\epsilon$ ($\epsilon>0$)
represents an infinitesimal imaginary part specifying on which side of
the cut a multi-valued function has to be evaluated. The definition of
the momentum-space integrals and of the arguments of the standard functions 
$B_0$, $C_0$, $D_0$ can be found in the appendix of \citere{di94}.

As an additional check 
of our one-loop results, we have analytically
compared the relative corrections $\M_1/\M_0$ with the corresponding
results $f^{(4)}/f^{(2)}$ of Milton et al.\ \cite{mi72}. Apart from a
trivially missing global sign in \citere{mi72} for the configuration
$({-}{+}{+}{+})$, which has also been pointed out by Swartz \cite{sw97},
we find complete agreement.

{}Finally, we consider the one-loop corrections to low-energy Compton
scattering in more detail. As already mentioned in the introduction, the
complete QED corrections to the cross sections
vanish in the low-energy limit $\beta\to 0$
owing to Thirring's theorem \cite{th50,di97}. Upon inspecting the
amplitudes and the phase space for photon bremsstrahlung, one finds that
all real ${\cal O}(\alpha)$ corrections vanish at least like $\beta^2$
relative to the Born cross section. Thus, only loop corrections contribute
in order $\be\al$ relative to the  lowest-order cross section. 
We have explicitly
expanded our results \refeq{eq:amponel} for $\beta\to 0$ and get
\beq
\M_1 = [\delta_{\mathrm{low}}+\O(\beta^2)\alpha] \, \M_0
\eeq
with the simple factors
\beqar\label{deltalow}
\delta_{\mathrm{low}}({\pm}{\pm}{\pm}{\pm}) &=& 
\frac{\alpha\beta}{\pi}\frac{\sin^2\big(\text\frac{\theta}{2}\big)}
{\cos^2\big(\text\frac{\theta}{2}\big)
 +\beta [2-\cos^2\big(\text\frac{\theta}{2}\big)]},
\nn\\[.5em]
\delta_{\mathrm{low}}({\pm}{\mp}{\pm}{\mp}) &=& 0,
\nn\\[.5em]
\delta_{\mathrm{low}}({\pm}{\mp}{\mp}{\pm}) &=& \frac{\alpha\beta}{\pi},
\nn\\[.5em]
\delta_{\mathrm{low}}({\pm}{\mp}{\mp}{\mp}) &=& 
\delta_{\mathrm{low}}({\pm}{\pm}{\mp}{\pm}) = -\frac{\alpha\beta}{2\pi}, 
\nn\\[.5em]
\delta_{\mathrm{low}}({\pm}{\pm}{\pm}{\mp}) &=& 
\delta_{\mathrm{low}}({\pm}{\mp}{\pm}{\pm}) = \frac{\alpha\beta}{2\pi},
\nn\\[.5em]
\delta_{\mathrm{low}}({\pm}{\pm}{\mp}{\mp}) &=& 
-\frac{\alpha\beta}{\pi}\sin^{-2}\big(\text\frac{\theta}{2}\big).
\eeqar
In the denominator of $\delta_{\mathrm{low}}({\pm}{\pm}{\pm}{\pm})$ we
keep the term $\cos^2\big(\text\frac{\theta}{2}\big)
+\beta [2-\cos^2\big(\text\frac{\theta}{2}\big)]$ exactly,
in order to avoid a mismatch when the pole in $\theta$ is cancelled 
in the product $\delta_{\mathrm{low}}\M_0$.

Using these expressions to calculate the corrected cross sections 
$\sigma(P_\Pe,P_\gamma)$,
and retaining only terms linear in $\be$, we find 
\beq
\frac{\rd\sigma(+1,\pm 1)}{\rd\cos\theta} =
[1+\delta_{\mathrm{low}}^\pm +\O(\beta^2)\alpha] \,
\frac{\rd\sigma_0(+1,\pm 1)}{\rd\cos\theta}, 
\qquad
\delta_{\mathrm{low}}^\pm = \mp \frac{\alpha\beta}{\pi}
\sin^2\big(\text\frac{\theta}{2}\big)
\disp\frac{1-3\cos\theta}{1+\cos^2\theta}.
\eeq
Owing to
\beq
\frac{\rd\sigma_0(+1,\pm 1)}{\rd\cos\theta} = 
\frac{\pi\alpha^2}{\Me^2}[1+\cos^2\theta + {\cal O}(\beta) ],
\label{eq:dsbeta0}
\eeq
this implies that there is no correction of order $\be\al$ 
to the unpolarized cross section in the low-energy limit. From 
\refeq{eq:dsbeta0} we can also deduce that the 
lowest-order asymmetry $A_{\mathrm{LR},0}$ 
vanishes like ${\cal O}(\beta)$. Thus, the 
relative correction $\delta_{\mathrm{A}}$ to $A_{\mathrm{LR},0}$
does not tend to zero for $\beta\to 0$, since the factors $\beta$ in
$\delta_{\mathrm{low}}^\pm$ are cancelled by a factor $\beta$ in 
$A_{\mathrm{LR},0}$. Explicitly we get
\beq\label{deltaAlow}
A_{\mathrm{LR}} = [1+\delta_{\mathrm{A,low}}+{\cal O}(\beta)\alpha] \,
A_{\mathrm{LR},0},
\qquad
\delta_{\mathrm{A,low}} = \frac{\alpha}{\pi}
\frac{3\cos\theta-1}{4(\beta+\cos\theta)},
\eeq
where again $\beta+\cos\theta$ in the denominator of
$\delta_{\mathrm{A,low}}$ was kept exactly, in order to account for the
exact pole position in $A_{\mathrm{LR},0}$.
The approximation \refeq{deltaAlow} is applicable to all asymmetries 
defined via differential cross sections that are linear distributions in
$\cos\theta$. Since
$\rd\bar E'_\Pe=\rd\bar E'_\gamma=E_\gamma\gab\beb\rd\cos\theta$
[see \refeq{eq:thboost}], the approximation is, in particular, valid for 
distributions in $\bar E'_\Pe$ and $\bar E'_\gamma$. The analogous
approximation for the correction to the asymmetry defined via total
cross sections can be obtained in a similar way; the result is
\beq
\delta_{\mathrm{A,low}} = \frac{3\alpha}{2\pi}.
\label{deltaAintlow}
\eeq

\section{Real corrections}
\label{se:real}

\subsection{Soft-photonic bremsstrahlung}

Radiative corrections resulting from the emission of soft photons are
proportional to the lowest-order cross section of the corresponding
process. The calculation of the relative correction factor
$\delta_{\mathrm{soft}}$ in ${\cal O}(\alpha)$ is straightforward
\cite{ADHab} and involves the integration over the phase space of the
emitted photon, the energy of which is bounded by the small soft-photon 
cut $\Delta E$. Note that $\delta_{\mathrm{soft}}$ is frame-dependent
owing to this condition. In the CM system the soft-photon correction
factor is given by
\beqar
\delta_{\mathrm{soft}} &=& \frac{\alpha}{\pi} \biggl\{
\ln\biggl(\frac{\lambda^2}{4\Delta E^2}\biggr)
+\frac{s+\Me^2}{s-\Me^2}\ln\biggl(\frac{s}{\Me^2}\biggr)
+\frac{t-2\Me^2}{\beta_t t}\biggl[ \,
        \ln(x_t)\ln\biggl(\frac{\lambda^2}{4\Delta E^2}\biggr)
        -\frac{1}{2}\ln^2\biggl(\frac{s}{\Me^2}\biggr)
\nn\\ && {} \quad
        -\Li\biggl(1-\frac{s}{\Me^2}x_t\biggr)
        -\Li\biggl(1-\frac{\Me^2}{s}x_t\biggr)
        +2\Re\{\Li(1+x_t)\}-\frac{\pi^2}{2}
\, \biggr] \biggr\},
\eeqar
where the $\ln\lambda^2$ terms cancel against the corresponding
IR-divergent contributions of the virtual corrections. In practice,
$\lambda$ can be set to any value providing a check of IR finiteness in
the sum of virtual and soft-photonic corrections. The dependence on the
soft-photon cut $\Delta E\ll E_\gamma$ drops out after adding the 
numerically integrated corrections induced by the process 
$\Pem\gamma\to\Pem\gamma\gamma$, where either emitted photon has 
an energy larger than $\Delta E$.

\subsection{Hard photon emission---the process
\boldmath{$\Pem\gamma\to\Pem\gamma\gamma$}}

The emission of an additional photon with finite energy leads us to the
kinematically different process
\beq
\Pem(p,\sigma) + \gamma(k,\lambda) \; \longrightarrow \;
\Pem(p',\sigma') + \gamma(k'_1,\lambda'_1) + \gamma(k'_2,\lambda'_2).
\eeq
The incoming momenta $p$, $k$ 
and the polarization vectors of the incoming photon 
are the same as in \refse{se:locs}. 
In the CM system the outgoing momenta $p'$ and $k'_n$ ($n=1,2$)
are specified by
\beqar
p^{\prime\mu} &=& E'_\Pe(1,\beta'\sin\theta'_\Pe\cos\phi'_\Pe,
\beta'\sin\theta'_\Pe\sin\phi'_\Pe,\beta'\cos\theta'_\Pe),
\qquad
\beta'=\sqrt{1-\Me^2/E_\Pe^{\prime2}},
\nn\\
k_n^{\prime\mu} &=&  
E'_{\gamma,n}(1,\sin\theta'_{\gamma,n}\cos\phi'_{\gamma,n},
\sin\theta'_{\gamma,n}\sin\phi'_{\gamma,n},\cos\theta'_{\gamma,n}).
\label{eq:eaeaa_mom}
\eeqar
The polarization vectors of the outgoing photons read 
\beqar
\veps^{\prime*\mu}(k'_n,\lambda'_n=\pm 1) &=& 
\frac{\re^{\pm\ri\phi'_{\gamma,n}}}{\sqrt{2}}(0,
-\cos\theta'_{\gamma,n}\cos\phi'_{\gamma,n}\mp \ri\sin\phi'_{\gamma,n},
\nl &&\qquad\quad\;
-\cos\theta'_{\gamma,n}\sin\phi'_{\gamma,n}\pm \ri\cos\phi'_{\gamma,n},
\sin\theta'_{\gamma,n}).
\label{eq:poldef2}
\eeqar
One of the angles $\phi'_{\gamma,n}$ or $\phi'_\Pe$ can be put to
zero upon orienting the coordinate system appropriately.
The lowest-order cross section for $\Pem\gamma\to\Pem\gamma\gamma$,
which yields an ${\cal O}(\alpha)$ correction to the Compton process, 
is given by 
\beq
\sigma_\gamma(P_\Pe,P_\gamma) = 
\frac{1}{4E_\gamma\sqrt{s}} \int \rd\Gamma_\gamma \,
\sum_{\sigma,\lambda,\sigma',\lambda'_1,\lambda'_2}
\frac{1}{8}(1+2\sigma P_\Pe)(1+\lambda P_\gamma) \,
|\M_\gamma(\sigma,\lambda,\sigma',\lambda'_1,\lambda'_2)|^2,
\label{eq:hbcs}
\eeq
where the phase-space integral is defined by
\beq
\int \rd\Gamma_\gamma = 
\int\frac{\rd^3 {\bf p}'}{(2\pi)^3 2E'_\Pe}
\int\frac{\rd^3 {\bf k}'_1}{(2\pi)^3 2E'_{\gamma,1}}
\int\frac{\rd^3 {\bf k}'_2}{(2\pi)^3 2E'_{\gamma,2}} \,
(2\pi)^4 \delta(p+k-p'-k'_1-k'_2).
\eeq
Since we sum over all polarizations in the final state and integrate
over the full three-particle phase space, we have included a
factor 1/2 in \refeq{eq:hbcs}, in order to compensate for double
counting of identical configurations of the photons. The phase-space
integration is performed numerically by 
{\sl Vegas} \cite{le78}.

\begin{figure}
\centerline{
\setlength{\unitlength}{1cm}
\begin{picture}(14,6.5)
\put(-2.5,-9.3){\includegraphics{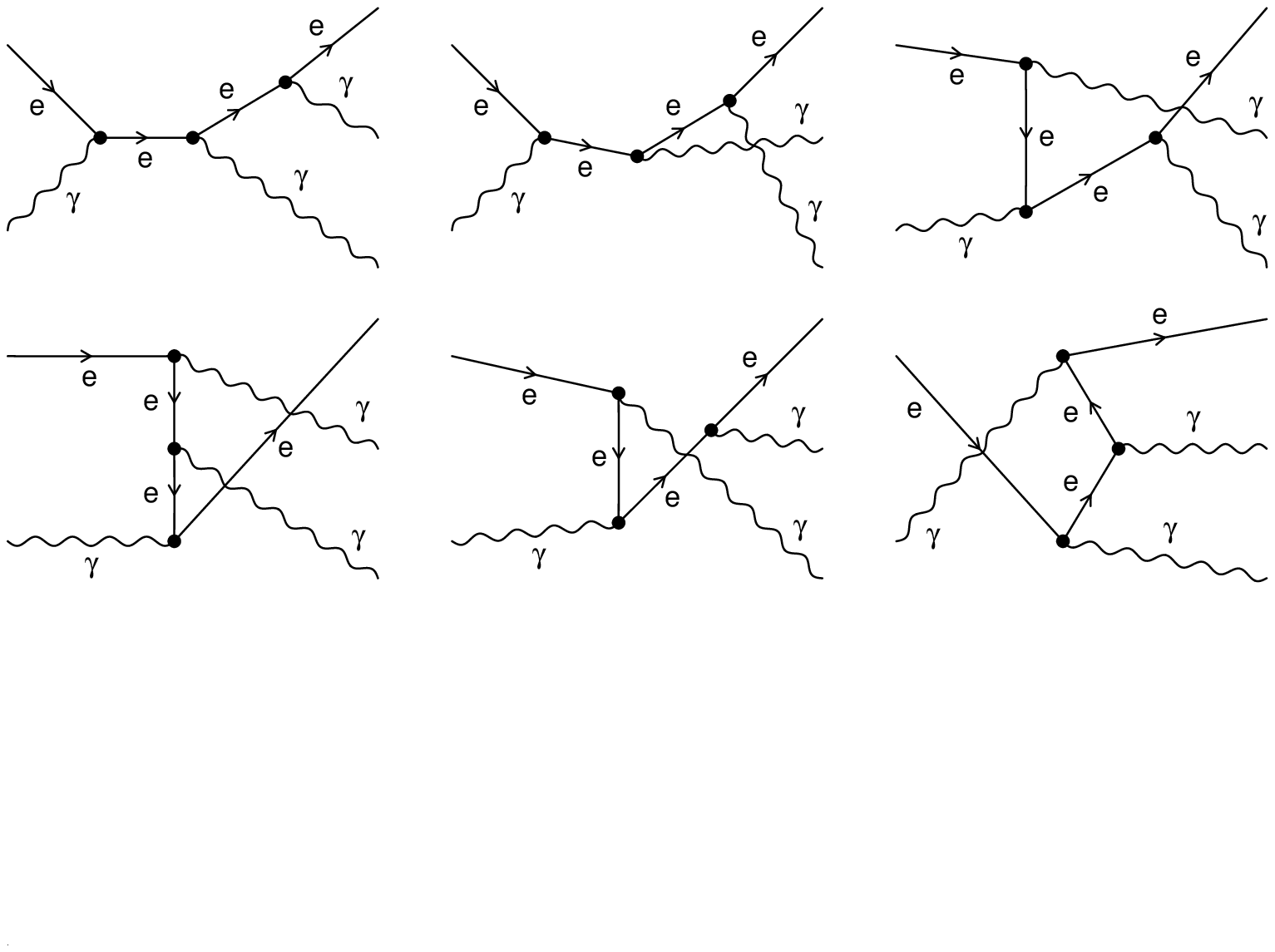}}
\end{picture} } 
\caption{Tree diagrams for $\Pem\gamma\to\Pem\gamma\gamma$}
\label{fi:eaeaadiags}
\end{figure}
The helicity amplitudes $\M_\gamma$, which receive contributions from the six
diagrams shown in \reffi{fi:eaeaadiags}, have been calculated in three 
independent ways, the results of which are in complete mutual numerical 
agreement. One calculation is based on the spinor calculus described in 
\citere{kl85}; 
the second one makes use of 
an explicit representation of spinors, polarization vectors, and Dirac
matrices. In the following we present the result of the third calculation,
which has been performed using the Weyl--van-der-Waerden (WvdW) formalism.
An introduction into this formalism and more details can be found in
\citere{di98}. In particular, the explicit form of the electron spinors,
which is needed for the transition to transversely polarized electrons,
can be found there. Here we give only the ingredients that are necessary
for the evaluation of the helicity amplitudes.

In the WvdW formalism, spinor chains and Minkowski inner products are
expressed in terms of antisymmetric spinor products
\beq
\langle\psi\phi\rangle = -\langle\phi\psi\rangle = 
\psi_1 \phi_2 - \psi_2 \phi_1
\eeq
of two-component WvdW spinors $\psi_A$, $\phi_A$. For each light-like
momentum there is an associated momentum spinor. Specifically,
we have the spinors
\beqar
k_A = \sqrt{2E_\gamma} \pmatrix{0 \cr 1}, 
\qquad
k'_{n,A} = \sqrt{2E'_{\gamma,n}} 
\pmatrix{ \Pe^{-\ri\phi'_{\gamma,n}}\cos\text\frac{\theta'_{\gamma,n}}{2} \cr
          \sin\text\frac{\theta'_{\gamma,n}}{2} }, 
\quad n=1,2,
\label{eq:kspinor}
\eeqar
for the photon momenta $k$ and $k'_n$ ($n=1,2$). For each time-like 
momentum there are two associated WvdW spinors; for $p$ we have
\beq
\kappa_{1,A} = \sqrt{E_\Pe(1+\beta)} \pmatrix{1 \cr 0}, \qquad
\kappa_{2,A} = \frac{\Me}{\sqrt{E_\Pe(1+\beta)}} \pmatrix{0 \cr -1}, 
\label{eq:pspinor}
\eeq
and for $p'$ 
\beq
\kappa'_{1,A} = \sqrt{E'_\Pe(1+\beta')}
\pmatrix{ \Pe^{-\ri\phi'_\Pe}\cos\text\frac{\theta'_\Pe}{2} \cr
          \sin\text\frac{\theta'_\Pe}{2} }, \qquad
\kappa'_{2,A} = \frac{\Me}{\sqrt{E'_\Pe(1+\beta')}}
\pmatrix{ \sin\text\frac{\theta'_\Pe}{2} \cr
          -\Pe^{+\ri\phi'_\Pe}\cos\text\frac{\theta'_\Pe}{2} }. 
\eeq
We note that 
$\langle\kappa_2\kappa_1\rangle=\langle\kappa'_2\kappa'_1\rangle=\Me$.
The different helicity amplitudes for $\Pem\gamma\to\Pem\gamma\gamma$ can 
be completely expressed in terms of spinor products of the WvdW spinors 
$k_A$, $k'_{1,A}$, $k'_{2,A}$, $\kappa_{1,A}$, $\kappa_{2,A}$, 
$\kappa'_{1,A}$, $\kappa'_{2,A}$. 
However, at some places it is convenient to keep the usual Minkowski inner 
products like $p\cdot k=p_\mu k^\mu$, etc. 
Moreover, we introduce the following abbreviations:
\beqar
\langle aPb\rangle &=& \sum_{i=1,2} 
\langle a\kappa_i\rangle^*\langle b\kappa_i\rangle, 
\nn\\
\langle aPP'a\rangle &=& \sum_{i,j=1,2} 
\langle a\kappa_i\rangle^*\langle\kappa'_j\kappa_i\rangle
\langle\kappa'_j a\rangle^*,
\qquad a,b = k,k'_1,k'_2.
\label{eq:spabb}
\eeqar

The presentation of the amplitudes can be shortened by exploiting
discrete symmetries. Helicity amplitudes with opposite helicity 
configurations are related by parity, leading to%
\footnote{The global signs in \refeq{parityrelation} and in similar
    relations below are convention-dependent. We consequently stick to the
    conventions of \citere{di98}.}
\beq\label{parityrelation}
\M_\gamma(-\sigma,-\lambda,-\sigma',-\lambda'_1,-\lambda'_2) =
\mathrm{sgn}(\sigma\sigma')
\M_\gamma(\sigma,\lambda,\sigma',\lambda'_1,\lambda'_2)^*,
\eeq
and the two outgoing photons can be interchanged owing to Bose symmetry,
\beq
\M_\gamma(\sigma,\lambda,\sigma',\lambda'_2,\lambda'_1) =
\M_\gamma(\sigma,\lambda,\sigma',\lambda'_1,\lambda'_2)
\Big|_{k'_1\leftrightarrow k'_2}.
\eeq
Making use of these relations, we get three 
independent amplitudes for fixed electron helicities: 
\newcommand{\MPK}{(p\cdot k)}
\newcommand{\MPKi}{(p\cdot k'_1)}
\newcommand{\MPKii}{(p\cdot k'_2)}
\newcommand{\MPpK}{(p'\cdot k)}
\newcommand{\MPpKi}{(p'\cdot k'_1)}
\newcommand{\MPpKii}{(p'\cdot k'_2)}
\newcommand{\KiKii}   {\langle k'_1 k'_2 \rangle}
\newcommand{\KiK}     {\langle k'_1 k \rangle}
\newcommand{\KiiK}    {\langle k'_2 k \rangle}
\newcommand{\CKiKii}  {\KiKii^*}
\newcommand{\CKiK}    {\KiK^*}
\newcommand{\CKiiK}   {\KiiK^*}
\newcommand{\KiPKii}  {\langle k'_1 P k'_2 \rangle}
\newcommand{\KiPK} {\langle k'_1 P k \rangle}
\newcommand{\KiiPK}{\langle k'_2 P k \rangle}
\newcommand{\KiiPKi}  {\langle k'_2 P k'_1 \rangle}
\newcommand{\KPKi} {\langle k P k'_1 \rangle}
\newcommand{\KPKii}{\langle k P k'_2 \rangle}
\newcommand{\KiPPpKi}     {\langle k'_1 P P' k'_1 \rangle}
\newcommand{\KiiPPpKii}   {\langle k'_2 P P' k'_2 \rangle}
\newcommand{\KPPpK} {\langle k P P' k \rangle}
\newcommand{\CKiPPpKi}    {\KiPPpKi^*}
\newcommand{\CKiiPPpKii}  {\KiiPPpKii^*}
\newcommand{\CKPPpK}{\KPPpK^*}
\newcommand{\Kiphi}   {\langle k'_1 \phi \rangle}
\newcommand{\Kiiphi}  {\langle k'_2 \phi \rangle}
\newcommand{\Kphi} {\langle k \phi \rangle}
\newcommand{\Kipsi}   {\langle k'_1 \psi \rangle}
\newcommand{\Kiipsi}  {\langle k'_2 \psi \rangle}
\newcommand{\Kpsi} {\langle k \psi \rangle}
\newcommand{\KiPHI}   {\langle k'_1 \phi' \rangle}
\newcommand{\KiiPHI}  {\langle k'_2 \phi' \rangle}
\newcommand{\KPHI} {\langle k \phi' \rangle}
\newcommand{\KiPSI}   {\langle k'_1 \psi' \rangle}
\newcommand{\KiiPSI}  {\langle k'_2 \psi' \rangle}
\newcommand{\KPSI} {\langle k \psi' \rangle}
\newcommand{\CKiphi}   {\Kiphi^*}
\newcommand{\CKiiphi}  {\Kiiphi^*}
\newcommand{\CKphi} {\Kphi^*}
\newcommand{\CKipsi}   {\Kipsi^*}
\newcommand{\CKiipsi}  {\Kiipsi^*}
\newcommand{\CKpsi} {\Kpsi^*}
\newcommand{\CKiPHI}   {\KiPHI^*}
\newcommand{\CKiiPHI}  {\KiiPHI^*}
\newcommand{\CKPHI} {\KPHI^*}
\newcommand{\CKiPSI}   {\KiPSI^*}
\newcommand{\CKiiPSI}  {\KiiPSI^*}
\newcommand{\CKPSI} {\KPSI^*}
\newcommand{\phiPSI}{\langle\phi\psi'\rangle}
\newcommand{\CphiPSI}{\phiPSI^*}
\beq
\M_\gamma(\sigma,\lambda,\sigma',\lambda'_1,\lambda'_2) =
\frac{e^3A_{\lambda\lambda'_1\lambda'_2}(\sigma,\sigma')}
{4\sqrt{2}\MPK\MPKi\MPKii}, 
\eeq
where
\beqar
A_{{+}{+}{+}}(\sigma,\sigma') &=& \Me\frac{(\CKiKii)^2}{2\MPpK}
\Big( -\CKPPpK\phiPSI+2\MPK\KPSI\Kphi \Big)
\nn\\
&& {} -\frac{\KiiPK}{\MPpKi} \Big[
\KiPPpKi \Big( \CKiipsi\KPSI+\CKiiPHI\Kphi \Big)
\nn\\
&& {} \phantom{ -\frac{\KiiPK}{\MPpKi} \Big[ }
+2\MPKi\Kphi\CKiPHI\CKiKii \Big]
\;+\; (k'_1\leftrightarrow k'_2),
\nn\\[.5em]
A_{{-}{+}{+}}(\sigma,\sigma') &=& \Me\phiPSI \Biggl[ 
 \frac{(\CKiiK)^2}{\MPpKi} \KiPPpKi 
-\frac{(\CKiKii)^2}{2\MPpK} \KPPpK \Biggr]
\;+\; (k'_1\leftrightarrow k'_2),
\nn\\[.5em]
A_{{-}{-}{+}}(\sigma,\sigma') &=& \Me\frac{(\CKiiK)^2}{\MPpKi}
\Big( \CKiPPpKi\phiPSI+2\MPKi\KiPSI\Kiphi \Big)
\nn\\
&& {} -\frac{\KPKi}{\MPpKii} \Big[
\KiiPPpKii \Big( \CKpsi\KiPSI+\CKPHI\Kiphi \Big)
\nn\\
&& {} \phantom{ -\frac{\KPKi}{\MPpKii} \Big[ }
+2\MPKii\Kiphi\CKiiPHI\CKiiK \Big]
\nn\\
&& {} +\frac{\KiiPKi}{\MPpK} \Big[
\KPPpK \Big( \CKiipsi\KiPSI+\CKiiPHI\Kiphi \Big)
\nn\\
&& {} \phantom{ +\frac{\KiiPKi}{\MPpK} \Big[ }
+2\MPK\Kiphi\CKPHI\CKiiK \Big].
\eeqar
The auxiliary spinors $\phi$, $\psi$, $\phi'$, $\psi'$ contain the
information about the electron helicity configurations $(\sigma,\sigma')$.
Their actual insertions are
\beq
(\phi,\psi) = \Biggl\{ 
\barr{ll} (\kappa_1,-\kappa_2) & \quad \mbox{for} \; \sigma=+, \\
          (\kappa_2, \kappa_1) & \quad \mbox{for} \; \sigma=-, \earr 
\qquad
(\phi',\psi') = \Biggl\{ 
\barr{ll} (\kappa'_1,-\kappa'_2) & \quad \mbox{for} \; \sigma'=+, \\
          (\kappa'_2, \kappa'_1) & \quad \mbox{for} \; \sigma'=-. \earr 
\label{eq:epolspinors}
\eeq
Actually, the quantity 
$A_{{-}{-}{+}}$ is related to $A_{{+}{+}{+}}$ via crossing symmetry 
for the photons $\gamma(k,\lambda)$ and $\gamma(k'_1,\lambda'_1)$,
leaving only two independent amplitudes.

\subsection{Associated pair creation---the process 
\boldmath{$\Pem\gamma\to\Pem\Pem\Pep$}}

If only electrons are detected in the final state, the reaction
\beq
\Pem(p,\sigma) + \gamma(k,\lambda) \; \longrightarrow \;
\Pem(p'_1,\sigma'_1) + \Pem(p'_2,\sigma'_2) + \Pep(q,\tau)
\eeq
represents an additional background process.
The incoming momenta $p$ and $k$ are again the same as in \refse{se:locs}. 
In the CM system the outgoing momenta $p'_n$ ($n=1,2$) and
$q$ are specified by
\beqar
p^{\prime\mu}_n &=& E'_{\Pe,n}(1,
\beta'_{\Pe,n}\sin\theta'_{\Pe,n}\cos\phi'_{\Pe,n},
\beta'_{\Pe,n}\sin\theta'_{\Pe,n}\sin\phi'_{\Pe,n},
\beta'_{\Pe,n}\cos\theta'_{\Pe,n}),
\nn\\
q^\mu  &=&  E_+(1,\beta_+\sin\theta_+\cos\phi_+,
\beta_+\sin\theta_+\sin\phi_+,\beta_+\cos\theta_+),
\eeqar
with $\beta'_{\Pe,n}=\sqrt{1-\Me^2/E_{\Pe,n}^{\prime 2}}$ and
$\beta_+=\sqrt{1-\Me^2/E_+^2}$. 
The lowest-order cross section for $\Pem\gamma\to\Pem\Pem\Pep$
is obtained as 
\beq
\sigma_\Pe(P_\Pe,P_\gamma) = 
\frac{1}{4E_\gamma\sqrt{s}} \int \rd\Gamma_\Pe \,
\sum_{\sigma,\lambda,\sigma'_1,\sigma'_2,\tau}
\frac{1}{8}(1+2\sigma P_\Pe)(1+\lambda P_\gamma) \,
|\M_\Pe(\sigma,\lambda,\sigma'_1,\sigma'_2,\tau)|^2,
\label{eq:eeecs}
\eeq
where the phase-space integral reads
\beq
\int \rd\Gamma_\Pe = 
\int\frac{\rd^3 {\bf p}'_1}{(2\pi)^3 2E'_{\Pe,1}}
\int\frac{\rd^3 {\bf p}'_2}{(2\pi)^3 2E'_{\Pe,2}}
\int\frac{\rd^3 {\bf   q}}{(2\pi)^3 2E_+} \,
(2\pi)^4 \delta(p+k-p'_1-p'_2-q).
\eeq
Double counting of identical configurations of the electrons in the
final state is avoided by including a factor 1/2 in \refeq{eq:eeecs}.
The phase-space integration is again performed numerically by 
{\sl Vegas} \cite{le78}.

\begin{figure}
\centerline{
\setlength{\unitlength}{1cm}
\begin{picture}(14,5.0)
\put(-2.5,-10.9){\includegraphics{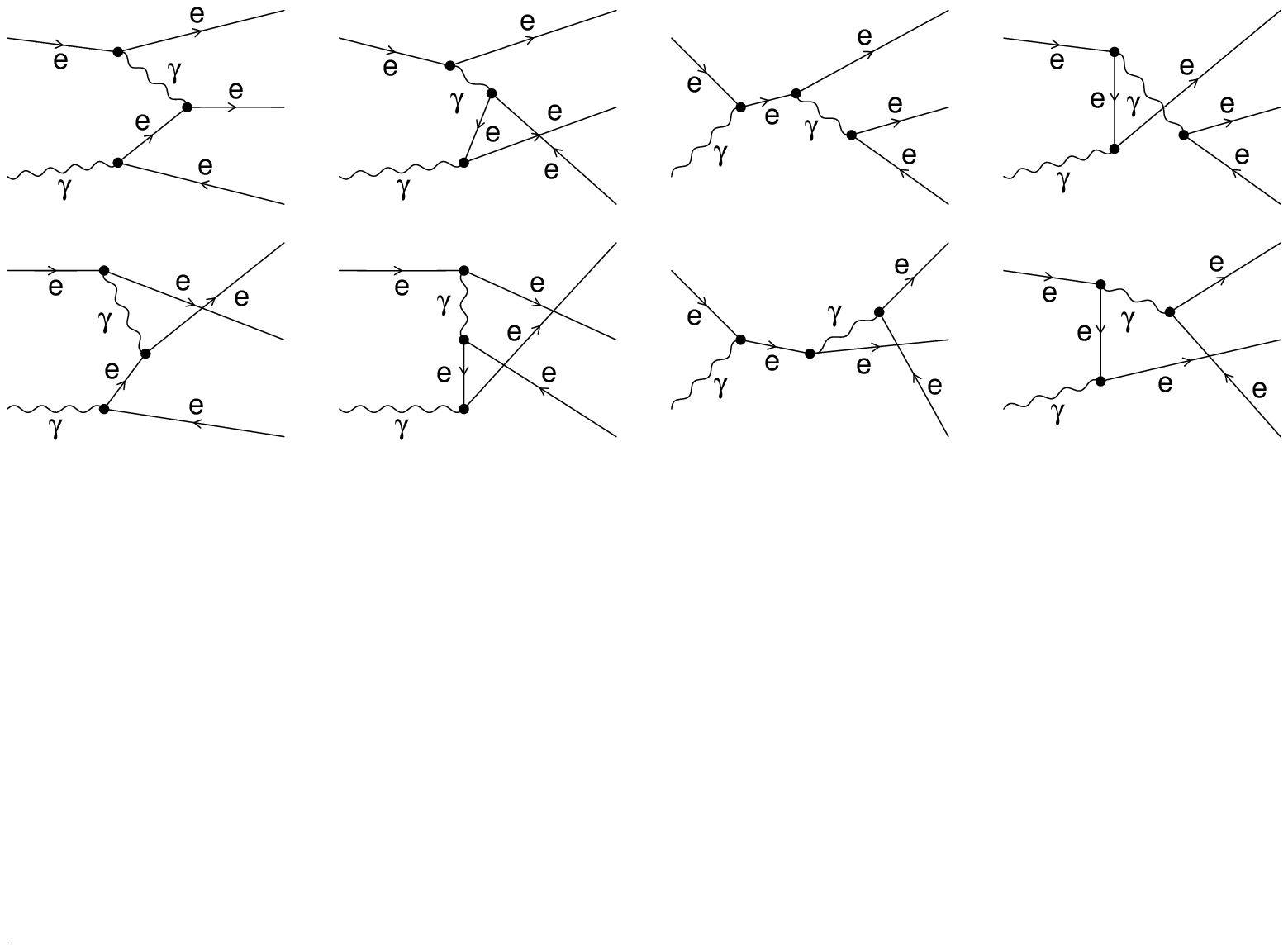}}
\end{picture} }
\caption{Tree diagrams for $\Pem\gamma\to\Pem\Pem\Pep$}
\label{fi:eaeeediags}
\end{figure}
The eight diagrams shown in \reffi{fi:eaeeediags} contribute to the
helicity amplitudes $\M_\Pe$ in lowest order. The 
amplitudes $\M_\Pe$ have been calculated in the WvdW
formalism as well as 
by using explicit spinors, polarization vectors, and Dirac matrices.
Here we again present
the result in terms of WvdW spinor products, as in the previous section
for $\Pem\gamma\to\Pem\gamma\gamma$. The spinors $\kappa_{i,A}$
($i=1,2$) and $k_A$ for the incoming momenta $p$ and $k$, 
are defined in \refeq{eq:kspinor} and \refeq{eq:pspinor}, respectively.
{}For the outgoing momenta $p'_n$ ($n=1,2$) and $q$ we introduce the following
spinors:
\beqar
\kappa'_{n1,A} &=& \sqrt{E'_{\Pe,n}(1+\beta'_{\Pe,n})}
\pmatrix{ \Pe^{-\ri\phi'_{\Pe,n}}\cos\text\frac{\theta'_{\Pe,n}}{2} \cr
          \sin\text\frac{\theta'_{\Pe,n}}{2} }, \quad
\kappa'_{n2,A} = \frac{\Me}{\sqrt{E'_{\Pe,n}(1+\beta'_{\Pe,n})}}
\pmatrix{ \sin\text\frac{\theta'_{\Pe,n}}{2} \cr
          -\Pe^{+\ri\phi'_{\Pe,n}}\cos\text\frac{\theta'_{\Pe,n}}{2} }, 
\nn\\
\rho_{1,A} &=& \sqrt{E_+(1+\beta_+)}
\pmatrix{ \Pe^{-\ri\phi_+}\cos\text\frac{\theta_+}{2} \cr
          \sin\text\frac{\theta_+}{2} }, \quad
\rho_{2,A} = \frac{\Me}{\sqrt{E_+(1+\beta_+)}}
\pmatrix{ \sin\text\frac{\theta_+}{2} \cr
          -\Pe^{+\ri\phi_+}\cos\text\frac{\theta_+}{2} }.
\hspace*{2em}
\eeqar
We note that 
$\langle\kappa'_{12}\kappa'_{11}\rangle=\langle\kappa'_{22}\kappa'_{21}\rangle
=\langle\rho_{2}\rho_{1}\rangle=\Me$. 
In analogy to \refeq{eq:spabb} we define
\beqar
\langle kPP'_n k\rangle &=& \sum_{i,j=1,2} 
\langle k\kappa_i\rangle^*\langle\kappa'_{nj}\kappa_i\rangle
\langle\kappa'_{nj} k\rangle^*,
\nn\\
\langle kQP'_n k\rangle &=& \sum_{i,j=1,2} 
\langle k\rho_i\rangle^*\langle\kappa'_{nj}\rho_i\rangle
\langle\kappa'_{nj} k\rangle^*,
\qquad n=1,2.
\eeqar

Helicity amplitudes with opposite helicity 
configurations are related by parity, yielding
\beq
\M_\Pe(-\sigma,-\lambda,-\sigma'_1,-\sigma'_2,-\tau) =
-\mathrm{sgn}(\sigma\sigma'_1\sigma'_2\tau)
\M_\Pe(\sigma,\lambda,\sigma'_1,\sigma'_2,\tau)^*,
\label{eq:eeeparity}
\eeq
and the amplitudes are antisymmetric with respect to the
interchange of the two outgoing electrons:
\beq
\M_\Pe(\sigma,\lambda,\sigma'_2,\sigma'_1,\tau) =
-\M_\Pe(\sigma,\lambda,\sigma'_1,\sigma'_2,\tau)
\Big|_{p'_1\leftrightarrow p'_2}.
\eeq

The actual calculation of the helicity amplitudes is most conveniently
performed by first considering the process $\Pem\gamma\to\Pem\Pf\bar\Pf$
with $\Pf\ne\Pem$, where only four diagrams contribute instead of eight.
The amplitudes for $\Pem\gamma\to\Pem\Pem\Pep$ are obtained from these 
intermediate results by first antisymmetrizing the amplitudes with
respect to the fields $\Pem$ and $\Pf$ in the final state, and then
setting $\Pf=\Pem$.
Using \refeq{eq:eeeparity}, all helicity amplitudes 
\beq
\M_\Pe(\sigma,\lambda,\sigma'_1,\sigma'_2,\tau) = 
-\frac{e^3}{\sqrt{2}} \biggl[A_{\lambda}(\sigma,\sigma'_1,\sigma'_2,\tau)
-A_{\lambda}(\sigma,\sigma'_2,\sigma'_1,\tau) \Big|_{p'_1\leftrightarrow p'_2, 
\phi'_1\leftrightarrow\phi'_2, \psi'_1\leftrightarrow\psi'_2} \biggr]
\eeq
can be deduced from one generic function
\newcommand{\MPiK} {(p'_1\cdot k)}
\newcommand{\MPiiK}{(p'_2\cdot k)}
\newcommand{\MPiiQ}{(p'_2\cdot q)}
\newcommand{\MQK}  {(q\cdot k)}
\newcommand{\MPPi} {(p\cdot p'_1)}
\newcommand{\KQPiiK}{\langle k Q P'_2 k \rangle}
\newcommand{\KPPiK} {\langle k P P'_1 k \rangle}
\newcommand{\CKQPiiK}{\KQPiiK^*}
\newcommand{\CKPPiK} {\KPPiK^*}
\newcommand{\KPSIi    }{\langle k \psi'_1 \rangle}
\newcommand{\Kxi      }{\langle k \xi \rangle}
\newcommand{\KPSIii   }{\langle k \psi'_2 \rangle}
\newcommand{\phixi    }{\langle \phi \xi \rangle}
\newcommand{\phiPSIii }{\langle \phi \psi'_2 \rangle}
\newcommand{\PHIieta  }{\langle \phi'_1 \eta \rangle}
\newcommand{\PHIiPHIii}{\langle \phi'_1 \phi'_2 \rangle}
\newcommand{\PHIiPSIii}{\langle \phi'_1 \psi'_2 \rangle}
\newcommand{\psieta   }{\langle \psi \eta \rangle}
\newcommand{\psiPHIii }{\langle \psi \phi'_2 \rangle}
\newcommand{\PSIixi   }{\langle \psi'_1 \xi \rangle}
\newcommand{\PSIiPSIii}{\langle \psi'_1 \psi'_2 \rangle}
\newcommand{\Cphixi    }{\phixi^*    }
\newcommand{\CphiPSIii }{\phiPSIii^* }
\newcommand{\CPHIieta  }{\PHIieta^*  }
\newcommand{\CPHIiPHIii}{\PHIiPHIii^*}
\newcommand{\CPHIiPSIii}{\PHIiPSIii^*}
\newcommand{\Cpsieta   }{\psieta^*   }
\newcommand{\CpsiPHIii }{\psiPHIii^* }
\newcommand{\CPSIixi   }{\PSIixi^*   }
\newcommand{\CPSIiPSIii}{\PSIiPSIii^*}
\beqar
A_+(\sigma,\sigma'_1,\sigma'_2,\tau) &=& 
\biggl\{ \frac{\CKQPiiK}{2\MPiiK\MQK[\Me^2-\MPPi]} 
       +\frac{\CKPPiK}{2\MPK\MPiK[\Me^2+\MPiiQ]} \biggr\}
\nn\\ &&{} \quad
\times \Big( \CpsiPHIii\PSIixi+\Cpsieta\PSIiPSIii+
                    \CPHIiPHIii\phixi+\CPHIieta\phiPSIii \Big)
\nn\\ &&{}
+\frac{1}{\Me^2-\MPPi} \biggl[ 
\frac{\Kxi}{\MQK}    \Big( \KPSIi\CpsiPHIii+\Kphi\CPHIiPHIii \Big)
\nn\\ &&{} 
\phantom{{}+\frac{1}{\Me^2-\MPPi} \biggl[}
-\frac{\KPSIii}{\MPiiK} \Big( \KPSIi\Cpsieta   +\Kphi\CPHIieta   \Big)
\biggr]
\nn\\ &&{}
+\frac{1}{\Me^2+\MPiiQ} \biggl[ 
\frac{\KPSIi}{\MPiK} \Big( \Kxi\CpsiPHIii+\KPSIii\Cpsieta \Big)
\nn\\ &&{} 
\phantom{{}+\frac{1}{\Me^2+\MPiiQ} \biggl[}
+\frac{\Kphi}{\MPK} \Big( \Kxi\CPHIiPHIii+\KPSIii\CPHIieta \Big)
\biggr],
\eeqar
where the spinors $\phi$, $\psi$ are the same as in \refeq{eq:epolspinors},
and $\phi'_n$, $\psi'_n$, $\xi$, $\eta$ ($n=1,2$) are given by
\beq
(\phi'_n,\psi'_n) = \Biggl\{ 
\barr{ll} (\kappa'_{n1},-\kappa'_{n2}) & \quad \mbox{for} \; \sigma'_n=+, \\
          (\kappa'_{n2}, \kappa'_{n1}) & \quad \mbox{for} \; \sigma'_n=-, \earr 
\qquad
(\xi,\eta) = \Biggl\{
\barr{ll} (-\rho_2, \rho_1) & \quad \mbox{for} \; \tau=+, \\
          ( \rho_1, \rho_2) & \quad \mbox{for} \; \tau=-. \earr
\eeq

The amplitude $A_+(\sigma,\sigma'_1,\sigma'_2,\tau)$
is composed of two gauge-invariant contributions, one of them consists
of all the terms involving the denominator $\Me^2-\MPPi$, the other
consists of the terms with the denominator $\Me^2+\MPiiQ$. 
The two gauge-invariant contributions correspond to the first two pairs of
graphs in \reffi{fi:eaeeediags} and are related to each other by 
appropriately interchanging the external fermions.

\section{Numerical results}
\label{se:numres}

In this section we apply our analytical results
to various cases of physical
relevance. We focus on polarimeters that determine the polarization of
electron beams by measuring the asymmetry \refeq{asymm}. The cross
sections entering this asymmetry are measured by either detecting the
scattered electron or the scattered photon. 
We consider, in particular, the polarimeters for CEBAF, 
for the SLC, and for an $\Pep\Pem$ collider with beam energies of $500\GeV$.

{}For the numerical evaluation we use 
\beq\label{params}
\alpha ^{-1}=137.0359895, \qquad \Me = 0.51099906 \MeV.
\eeq

\subsection{Results for the CEBAF polarimeter}

As a first application we discuss the CEBAF polarimeter
\cite{cebaf}. Here an electron beam with an energy of a few GeV 
collides with a laser beam of a few eV. We consider three setups:
\beq\label{cebaf}
\barr{c@{\quad}c@{\quad}|@{\quad}c@{\quad}c@{\quad}c@{\quad}c}
\bar E_{\ga}\ [\mathrm{eV}] & \bar E_{\Pe} \ [\mathrm{GeV}]  &
E_{\CM}\ [\mathrm{MeV}] &
\bar E'_{\gamma,\max}\ [\mathrm{GeV}] & \be & \gab/10^3 \\
\hline
 1.165 & 4.0 &  0.5289  &   0.2665 &  0.03446  & 7.563 \\
 1.165 & 6.0 &  0.5377  &   0.5803 &  0.05082  & 11.16 \\
 1.165 & 8.0 &  0.5463  &   0.9995 &  0.06663  & 14.65
\earr
\eeq
The backward-scattered photons are detected by a calorimeter 
that is located roughly $5\,$m after the interaction point perpendicular to
the incident electron with
transverse extension of $10\,$cm$\times10\,$cm. If two photons (from
double Compton scattering events) hit the detector,
only the sum of their energies is detected. 

We have implemented the detector in our calculation. The hard-photon
part of the differential cross section $\rd\si/\rd \bar E'_{\ga}$ is
obtained as the sum of the cross section where two photons hit the
detector with $\bar E'_{\ga}=\bar E'_{\ga,1}+\bar E'_{\ga,2}$ and the
cross sections where exactly one photon hits the detector with 
$\bar E'_{\ga}=\bar E'_{\ga,1}$ or $\bar E'_{\ga}=\bar E'_{\ga,2}$.
Since the final-state photons are 
strongly boosted in
the backward direction, almost all photons hit the detector.
Numerically our results with and without detector cuts practically
coincide, i.e.\ the finite extension of the detector turns out to be
irrelevant.

In \reftas{tab:ceb1}, \ref{tab:ceb2} and in
\reffis{fig:cebsigma}, \ref{fig:cebasymm} we show the 
unpolarized lowest-order cross section 
$\rd\si_0/\rd \bar E'_{\ga}$ and the lowest-order asymmetry 
$A_{\mathrm{LR,0}}$ 
as a function of the photon energy $\bar E'_{\ga}$ deposited in the
calorimeter. In addition, we show the corrections to these quantities.
\btab
$$
\newcommand{\m}{\phantom{-}}
\begin{array}{c@{\quad}c@{\quad}c@{\quad}c@{\quad}c@{\quad}c@{\quad}c}
\hline\hline
\rule[-1.5ex]{0ex}{3.5ex}
{\bar E'_{\ga} \ [\mathrm{GeV}]} &
{\dsidea \ [\mathrm{mb}/\mathrm{GeV}]}&
\de\  [\%] &
{\ \ A_{\mathrm{LR},0}\ \ } &
{\De A_{\mathrm{LR}}} \times 100 & 
\de_{\mathrm{A}}\  [\%] & \de_{\mathrm{A,low}}\  [\%] \\
\hline  
   0.010  &    3258.0  &  \m0.0038  &    -0.0025  &  -0.0001  & \m0.05 & \m0.11 \\
   0.035  &    2737.5  &  \m0.0037  &    -0.0084  &  -0.0007  & \m0.08 & \m0.09 \\
   0.060  &    2322.2  &  \m0.0035  &    -0.0130  &  -0.0008  & \m0.06 & \m0.06 \\
   0.085  &    2016.3  &  \m0.0033  &    -0.0146  &  -0.0003  & \m0.02 & \m0.01 \\
   0.110  &    1824.1  &  \m0.0030  &    -0.0111  & \m0.0012  &  -0.10 &  -0.13 \\
   0.135  &    1750.2  &  \m0.0025  &    -0.0015  & \m0.0036  &  -2.48 &  -2.82 \\
   0.160  &    1799.2  &  \m0.0021  &   \m0.0133  & \m0.0068  & \m0.51 & \m0.56 \\
   0.185  &    1976.0  &  \m0.0017  &   \m0.0300  & \m0.0099  & \m0.33 & \m0.36 \\
   0.210  &    2285.7  &  \m0.0013  &   \m0.0456  & \m0.0126  & \m0.28 & \m0.29 \\
   0.235  &    2733.4  &  \m0.0009  &   \m0.0580  & \m0.0144  & \m0.25 & \m0.26 \\
   0.260  &    3324.7  &  -0.0003  &    \m0.0670  & \m0.0156  & \m0.23 & \m0.24 \\
\hline\hline
\end{array}
$$
\caption{Lowest-order unpolarized cross section  and asymmetry
  together with the corresponding corrections for the CEBAF
  polarimeter   ($\bar E_\Pe=4\GeV$, $\bar E_\ga=1.165\eV$)}
\label{tab:ceb1}
\etab
\btab
$$
\newcommand{\m}{\phantom{-}}
\begin{array}{c@{\quad}c@{\quad}c@{\quad}c@{\quad}c@{\quad}c@{\quad}c}
\hline\hline
\rule[-1.5ex]{0ex}{3.5ex}
{\bar E'_{\ga} \ [\mathrm{GeV}]} &
{\dsidea \ [\mathrm{mb}/\mathrm{GeV}]}&
\de\  [\%] &
{\ \ A_{\mathrm{LR},0}\ \ } &
{\De A_{\mathrm{LR}}} \times 100 & 
\de_{\mathrm{A}} \  [\%] & \de_{{\mathrm{A,low}}}\  [\%] \\
\hline  
    0.010  &     858.50  & \m0.0111  &    -0.0013  & \m0.0007  &  -0.52 & \m0.11 \\ 
    0.100  &     732.56  & \m0.0107  &    -0.0123  &  -0.0005  & \m0.04 & \m0.09 \\ 
    0.200  &     616.50  & \m0.0103  &    -0.0230  &  -0.0012  & \m0.05 & \m0.07 \\ 
    0.300  &     527.60  & \m0.0098  &    -0.0288  &  -0.0008  & \m0.03 & \m0.02 \\ 
    0.400  &     468.14  & \m0.0091  &    -0.0252  & \m0.0013  &  -0.05 &  -0.09 \\ 
    0.500  &     440.57  & \m0.0080  &    -0.0085  & \m0.0057  &  -0.67 &  -0.88 \\ 
    0.600  &     447.56  & \m0.0068  &   \m0.0207  & \m0.0118  & \m0.57 & \m0.69 \\ 
    0.700  &     491.99  & \m0.0057  &   \m0.0559  & \m0.0183  & \m0.33 & \m0.38 \\ 
    0.800  &     576.99  & \m0.0046  &   \m0.0889  & \m0.0238  & \m0.27 & \m0.30 \\ 
    0.900  &     705.95  & \m0.0029  &   \m0.1149  & \m0.0277  & \m0.24 & \m0.27 \\ 
    0.990  &     862.64  &  -0.0036  &   \m0.1313  & \m0.0298  & \m0.23 & \m0.25 \\ 
\hline\hline
\end{array}
$$
\caption{Lowest-order unpolarized cross section  and asymmetry together
  with the corresponding corrections for the CEBAF polarimeter
  ($\bar E_\Pe=8\GeV$, $\bar E_\ga=1.165\eV$)}
\label{tab:ceb2}
\etab
\btab
$$
\begin{array}{c@{\quad}c@{\quad}c@{\quad}c@{\quad}c@{\quad}c@{\quad}c@{\quad}c}
\hline\hline
\rule[-1.5ex]{0ex}{3.5ex}
{\bar E_{\ga} \ [\mathrm{GeV}]} &
{\bar E_{\Pe} \ [\mathrm{GeV}]} &
{\si_0}\ [\mathrm{mb}]&
\de\  [\%] &
{\ \ A_{\mathrm{LR},0}\ \ } &
{\De A_{\mathrm{LR}}} \times 100 &
\de_{\mathrm{A}}\  [\%] &
\de_{\mathrm{A,low}}\  [\%] \\
\hline  
1.165 &   4.0  &    621.80  &  0.0023  &    0.0160  & 0.0057 & 0.35 & 0.35 \\
1.165 &   6.0  &    602.74  &  0.0045  &    0.0228  & 0.0082 & 0.36 & 0.35 \\
1.165 &   8.0  &    585.17  &  0.0071  &    0.0289  & 0.0104 & 0.36 & 0.35 \\
\hline\hline
\end{array}
$$
\caption{Integrated lowest-order unpolarized cross section and
  left--right asymmetry together with the corresponding corrections 
 for the CEBAF polarimeter}
\label{tab:cebi}
\etab
\bfi
\centerline{
\setlength{\unitlength}{1cm}
\begin{picture}(16,9)
\put(0,-1.5){\includegraphics{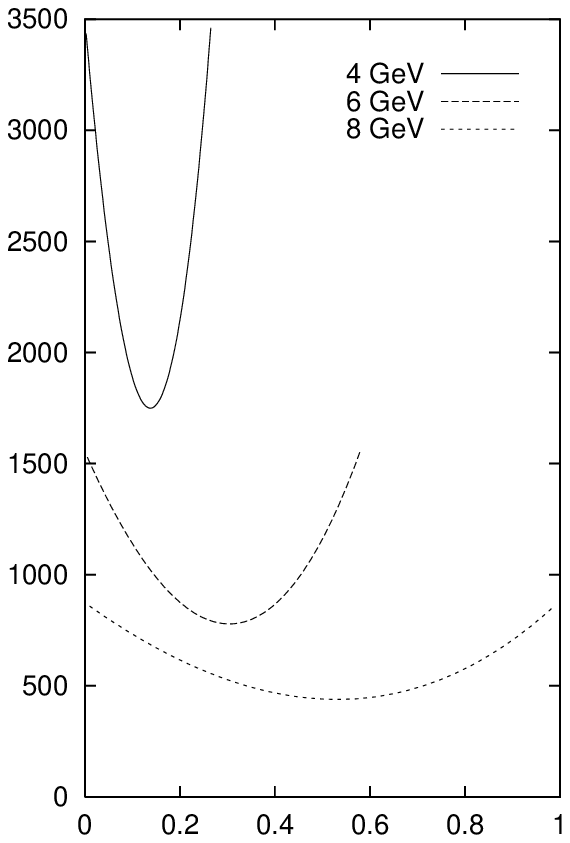}}
\put(8,-1.5){\includegraphics{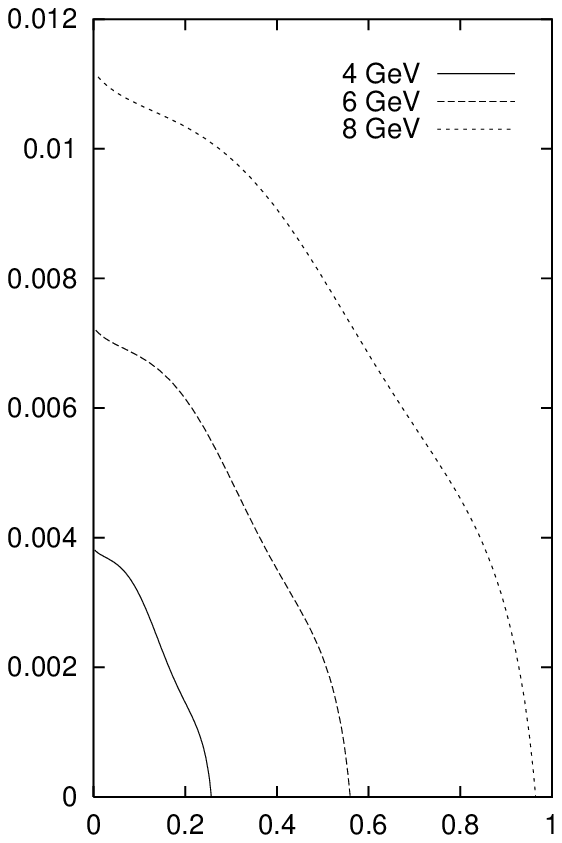}}
\put(0.5,5.5){\makebox(1,1)[c]{{$\disp\frac{\rd\si_0}{\rd\bar E'_{\ga}} $}}}
\put(0,4.5){\makebox(2,1)[c]{{$ [\mathrm{mb}/\mathrm{GeV}]$}}}
\put(9.3,5.0){\makebox(1,1)[c]{$\de\  [\%]$}}
\put(5.0,-0.3){\makebox(1,1)[cc]{$\bar E'_\ga/\mathrm{GeV}$}}
\put(13.0,-0.3){\makebox(1,1)[cc]{$\bar E'_\ga/\mathrm{GeV}$}}
\end{picture}
}
\caption[]{Lowest-order unpolarized cross section (left) and
  corresponding percentage
  relative corrections (right) for the CEBAF polarimeter 
  ($\bar E_\ga=1.165\eV$)}
\label{fig:cebsigma}
\vspace{2em}
\centerline{
\setlength{\unitlength}{1cm}
\begin{picture}(16,9)
\put(0,-1.5){\includegraphics{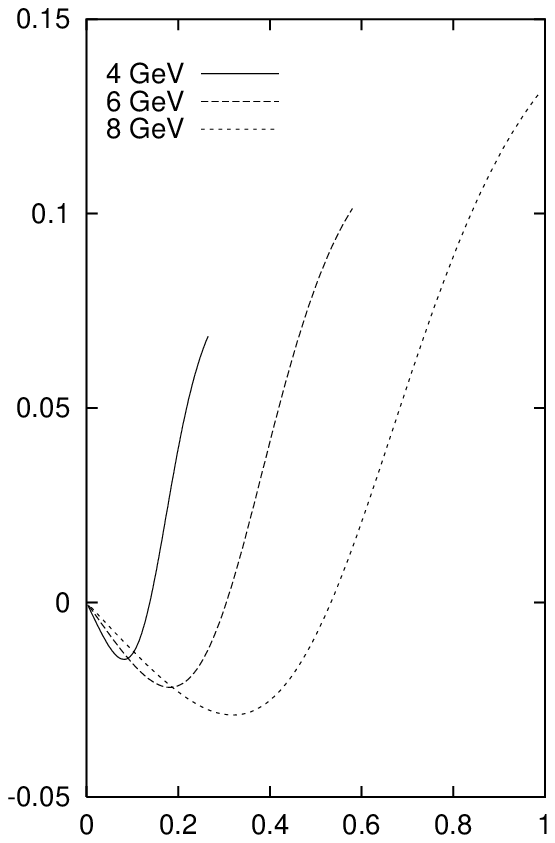}}
\put(8,-1.5){\includegraphics{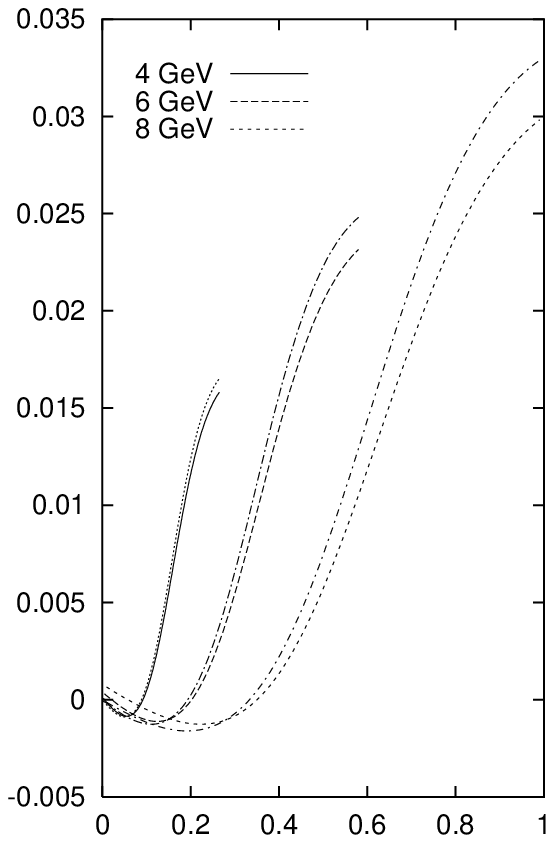}}
\put(1,5){\makebox(1,1)[c]{${A_{\mathrm{LR},0}}$}}
\put(9.0,5.0){\makebox(1,1)[c]{$100\times \De A_{\mathrm{LR}}$}}
\put(5,-0.3){\makebox(1,1)[cc]{$\bar E'_\ga/\mathrm{GeV}$}}
\put(13,-0.3){\makebox(1,1)[cc]{$\bar E'_\ga/\mathrm{GeV}$}}
\end{picture}
}
\caption[]{Lowest-order left--right asymmetry (left) and corresponding absolute
  corrections (right) for the CEBAF polarimeter ($\bar
  E_\ga=1.165\eV$). In the right plot, the lines next to the labelled
  ones correspond to the respective low-energy approximation.}
\label{fig:cebasymm}
\efi

The relative corrections $\de$ to the unpolarized cross section,
defined by
\beq
\frac{\rd\si}{\rd\bar E'_\ga} = \frac{\rd\si_0}{\rd\bar E'_\ga}(1+\de),
\eeq
are below roughly
$10^{-4}$, take their largest values for small photon energies
and decrease with increasing photon energy.  
The smallness of these corrections can be explained by the suppression with a factor
$\beta^2\lsim 4\times 10^{-3}$ (see \refse{se:virt}).
The absolute corrections $\De A_{\mathrm{LR}}$ to the asymmetry 
reach at most few $10^{-4}$ for large photon energies and
roughly follow the shape of the lowest-order asymmetries. 
The relative corrections $\de_{\mathrm{A}}$ to $A_{\mathrm{LR}}$,
defined by 
\beq 
A_{\mathrm{LR}} =
A_{\mathrm{LR,0}}(1+\de_{\mathrm{A}}), 
\eeq 
are of the expected order  
$\alpha/\pi\sim 2\times 10^{-3}$, as long as the
lowest-order asymmetry is not suppressed. Note that the low-energy
approximation $\de_{\mathrm{A,low}}$ for $\de_{\mathrm{A}}$, 
which is given in \refeq{deltaAlow}, works very well for CEBAF 
energies, where $\beta<0.07$.

The total cross sections and the related asymmetries are listed
together with the corresponding corrections in \refta{tab:cebi}.
The corresponding low-energy approximation $\de_{\mathrm{A,low}}$ 
is given in \refeq{deltaAintlow}.
If detector cuts on the outgoing photons are applied, the distribution 
$\rd\si/\rd\bar E'_\ga$ becomes IR-singular for $\bar E'_\ga\to 0$. 
For a well-defined total cross section
$\si$, either a lower cut $\De E_\ga$ for $\bar E'_\ga$ 
has to be introduced, or the detector cuts have to be dropped;
the latter was done for \refta{tab:cebi}.

In conclusion, the relative corrections to the polarization asymmetry
measured at \mbox{CEBAF} are typically ${\lsim 0.4\%}$, whenever the
asymmetry is sizeable.  For a polarization measurement at the $1\%$
level the simple low-energy approximation is therefore fully
sufficient.

\subsection{Results for the SLD polarimeter}

Next we consider the SLD polarimeter \cite{sld}. In this case an
electron beam of about $45\GeV$ collides with a 
$2.33\eV$ photon beam. 
The scattered electrons are detected, and their energy is measured.
Since we are going to perform a direct comparison of our results to the 
ones of Veltman \cite{ve89} and Swartz \cite{sw97}, we take the input values
\beqar
\bar E_{\ga} &=& 2.34\eV, \qquad \bar E_{\Pe} = 50\GeV,
\eeqar
leading to
\beq
E_{\CM} = 0.8539 \MeV, \qquad 
\bar E'_{\Pe,\min} =  17.91 \GeV, 
\qquad \be = 0.4726, \qquad\gab =5.856 \times 10^4.
\eeq

As for the CEBAF detector, the finite extension of the detector is
irrelevant because of the even stronger boost from the LAB
system to the CM system.

In \refta{tab:slc} we show the lowest-order cross section 
$\rd\si_0/\rd \bar E'_{\Pe}$ and the asymmetry defined in \refeq{asymm}
as a function of the energy of the scattered electron $\bar E'_{\Pe}$
together with the corresponding corrections. 
\btab
$$
\catcode`!=\active
\newcommand{!}{\phantom{-}}
\begin{array}{c@{\quad}c@{\quad}c@{\quad}c@{\quad}c@{\quad}c}
\hline\hline
\rule[-1.5ex]{0ex}{3.5ex}
{\bar E'_{\Pe} \ [\mathrm{GeV}]} &
{\dsidee \ [\mathrm{mb}/\mathrm{GeV}]}&
\de\  [\%] &
{A_{\mathrm{LR},0}} &
{\De A_{\mathrm{LR}}} \times 100 & 
\de_{\mathrm{A}}\  [\%] \\
\hline  
   17.92  &      17.504  &    -0.63  &    !0.7717  &     0.0725 & !0.09 \\
   19.90  &      13.272  &    !0.10  &    !0.6102  &     0.0572 & !0.09 \\
   21.90  &      10.621  &    !0.16  &    !0.4177  &     0.0478 & !0.11 \\
   23.90  &      ~9.008  &    !0.18  &    !0.2181  &     0.0383 & !0.18 \\
   25.90  &      ~8.073  &    !0.18  &    !0.0374  &     0.0293 & !0.78 \\
   27.90  &      ~7.592  &    !0.17  &    -0.1051  &     0.0226 &  -0.22 \\
   29.90  &      ~7.420  &    !0.17  &    -0.2014  &     0.0195 &  -0.10 \\
   31.90  &      ~7.461  &    !0.16  &    -0.2544  &     0.0199 &  -0.08 \\
   33.90  &      ~7.649  &    !0.16  &    -0.2726  &     0.0232 &  -0.09 \\
   35.90  &      ~7.941  &    !0.15  &    -0.2657  &     0.0284 &  -0.11 \\
   37.90  &      ~8.305  &    !0.15  &    -0.2422  &     0.0347 &  -0.14 \\
   39.90  &      ~8.719  &    !0.15  &    -0.2085  &     0.0418 &  -0.20 \\
   41.90  &      ~9.167  &    !0.15  &    -0.1693  &     0.0492 &  -0.29 \\
   43.90  &      ~9.637  &    !0.16  &    -0.1274  &     0.0568 &  -0.45 \\
   45.90  &      10.122  &    !0.17  &    -0.0848  &     0.0645 &  -0.76 \\
   47.90  &      10.614  &    !0.19  &    -0.0428  &     0.0723 &  -1.69 \\
   49.90  &      11.111  &    !0.22  &    -0.0020  &     0.0801 &  -40.0 \\
\hline\hline
\end{array}
$$
\caption{Lowest-order unpolarized cross section  and asymmetry
  together with the 
  corresponding corrections for the SLD polarimeter
  ($\bar E_\Pe=50\GeV$, $\bar E_\ga=2.34\eV$)}
\label{tab:slc}
\etab 
{}For SLC energies, the
corrections to the cross section are no longer suppressed by $\be$ and
are of the expected order $\alpha/\pi\sim 2\times
10^{-3}$.  The numerical analysis shows that the low-energy
approximation \refeq{deltaAlow} becomes useless and 
it is thus not given in the table.  

The results for the total unpolarized cross section are
\beq
\si_0 = 299.89\mba, \qquad \delta = 0.14\%, 
\eeq
and for the related asymmetry we get
\beq
A_{\mathrm{LR},0} = 0.03089, \qquad \De A_{\mathrm{LR}} =  0.00031,
\qquad \delta_{\mathrm{A}} = 1.0\%.
\eeq

\bfi
\centerline{
\setlength{\unitlength}{1cm}
\begin{picture}(16,9)
\put(0,-1.5){\includegraphics{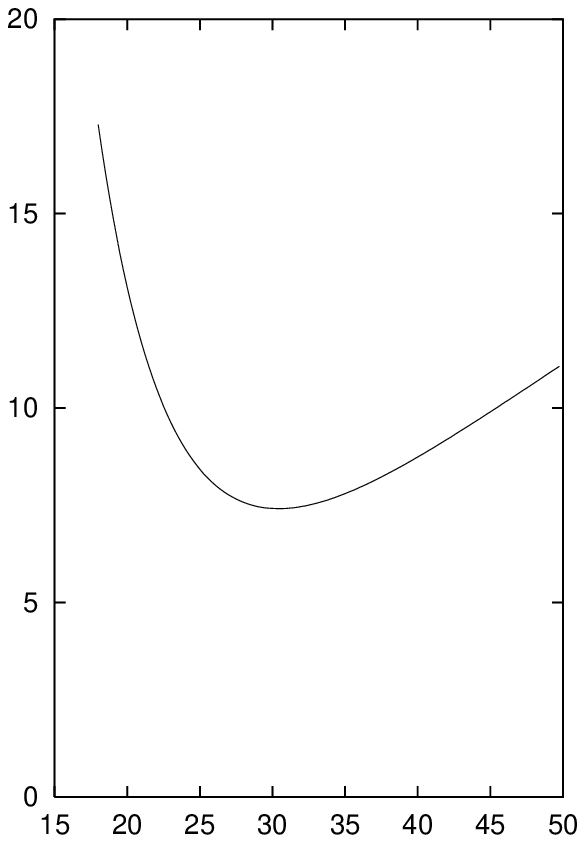}}
\put(8,-1.5){\includegraphics{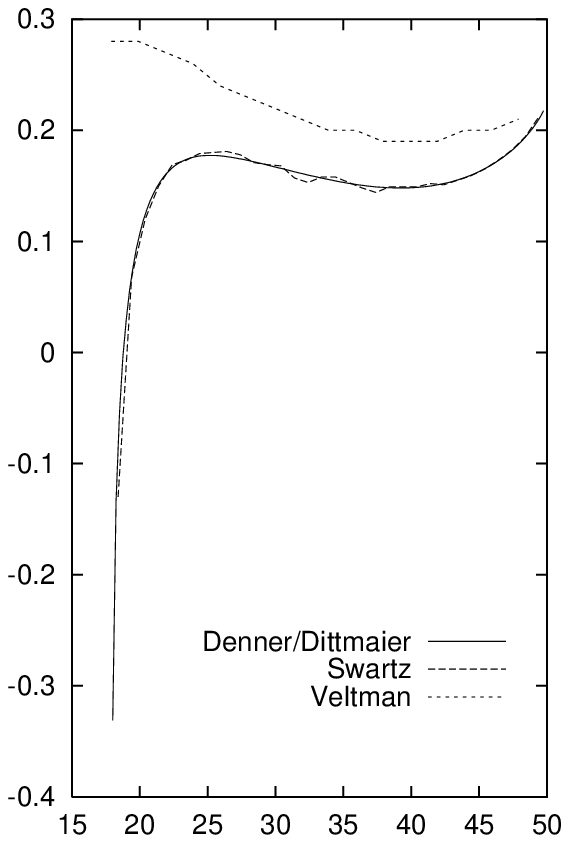}}
\put(0.5,5.5){\makebox(1,1)[c]{{$\disp\frac{\rd\si_0}{\rd\bar E'_{\Pe}} $}}}
\put(0,4.5){\makebox(2,1)[c]{{$ [\mathrm{mb}/\mathrm{GeV}]$}}}
\put(9.3,5.0){\makebox(1,1)[c]{$\de\  [\%]$}}
\put(5.0,-0.3){\makebox(1,1)[cc]{$\bar E'_\Pe/\mathrm{GeV}$}}
\put(13.0,-0.3){\makebox(1,1)[cc]{$\bar E'_\Pe/\mathrm{GeV}$}}
\end{picture}
}
\caption[]{Lowest-order unpolarized cross section (left) and
  corresponding percentage
  relative corrections (right) for the SLD polarimeter 
  ($\bar E_\Pe=50\GeV$, $\bar E_\ga=2.34\eV$)}
\label{fig:slcsigma}
\vspace{2em}
\centerline{
\setlength{\unitlength}{1cm}
\begin{picture}(16,9)
\put(0,-1.5){\includegraphics{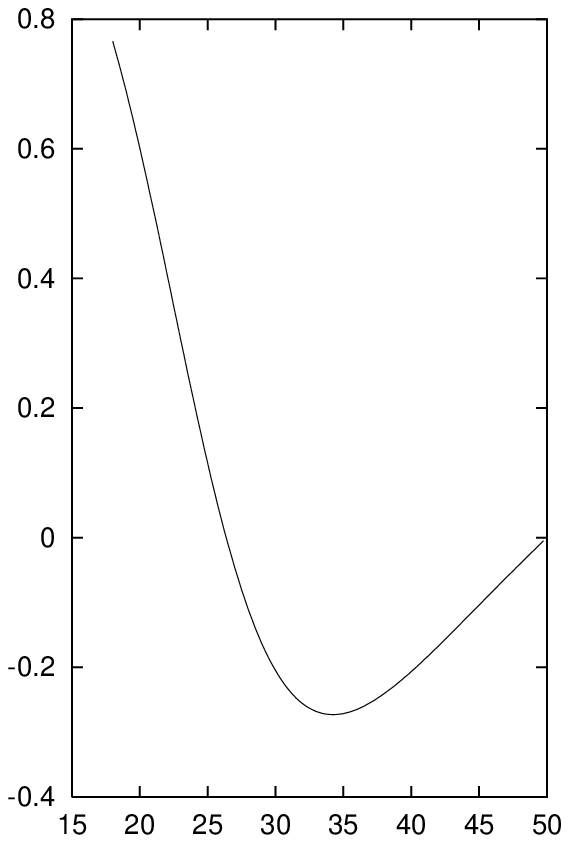}}
\put(8,-1.5){\includegraphics{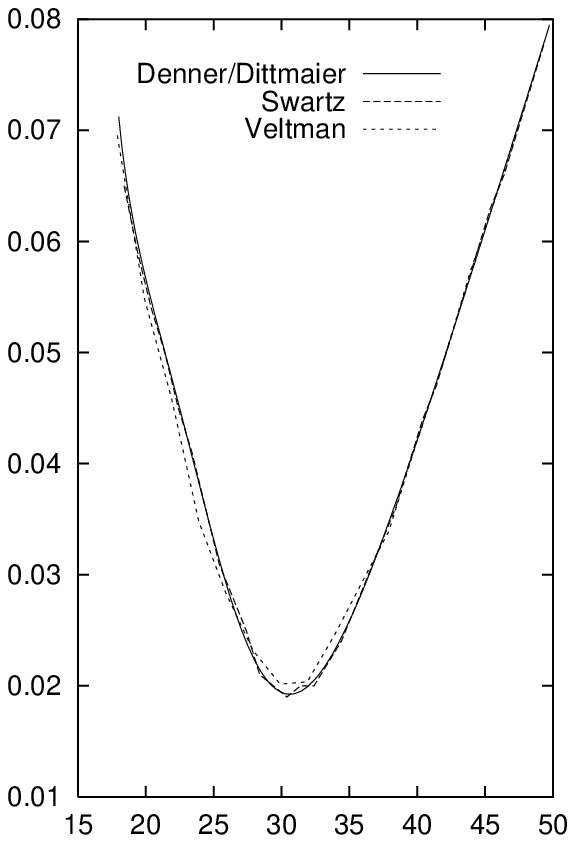}}
\put(1,5){\makebox(1,1)[c]{${A_{\mathrm{LR},0}}$}}
\put(9.0,5.4){\makebox(1,1)[c]{$100\times \De A_{\mathrm{LR}}$}}
\put(5,-0.3){\makebox(1,1)[cc]{$\bar E'_\Pe/\mathrm{GeV}$}}
\put(13,-0.3){\makebox(1,1)[cc]{$\bar E'_\Pe/\mathrm{GeV}$}}
\end{picture}
}
\caption[]{Lowest-order left--right asymmetry (left) and corresponding absolute
  corrections (right) for the SLD polarimeter 
  ($\bar E_\Pe=50\GeV$, $\bar E_\ga=2.34\eV$)}
\label{fig:slcasymm}
\efi
In \reffis{fig:slcsigma} and \ref{fig:slcasymm} we compare our 
results for the corrections to the unpolarized cross section and the
asymmetry with the results of Swartz\cite{sw97}, which we directly
obtained from the author, and Veltman, which are taken from
\citere{ve89}. For the corrections to the cross section, where the
results of Swartz and Veltman disagree, we reproduce the results of
Swartz well. For the asymmetry we confirm the results of Swartz who
has found small deviations from Veltman.

As pointed out by Swartz, the inclusion of the $\Oa$ corrections
increases the measured value of $A_{\mathrm{LR}}$ by $0.1\%$, 
since the polarization is 
determined using only scattered electrons with energies 
\mbox{$\lsim 20\GeV$}.
This effect is too small and has the wrong sign to account for the
discrepancy \cite{LEPEWWG97} in the experimental results for the
effective leptonic weak mixing angle between SLC and LEP.

\subsection{Results for an NLC polarimeter}

{}Finally, we consider a polarimeter for a NLC. 
We assume an ``SLD-like'' polarimeter with the parameters 
\beq
\bar E_{\ga} = 2.33\eV, \qquad \bar E_{\Pe} = 500\GeV,
\eeq
leading to
\beq
E_{\CM} =  2.218 \MeV, \qquad \bar E'_{\Pe,\min} =  26.53 \GeV,
\qquad \be = 0.8992, \qquad \gab=2.254 \times 10^5.
\eeq
In this case we are 
at the onset of the relativistic regime,
where because of enhanced logarithms 
the relative corrections start to exceed the naive expectation 
of the order $\alpha/\pi\sim 2\times 10^{-3}$.
Moreover, in addition to the
bremsstrahlung process $\Pem\ga\to\Pem\ga\ga$,
also the associated
pair-production process $\Pem\ga\to\Pem\Pem\Pep$ contributes in the
energy region
\beq
\bar E'_{\Pe,\min} =  34.36 \GeV <\bar E'_{\Pe} < 
\bar E'_{\Pe,\max} =  386.05 \GeV .
\eeq
We assume that both of the final-state electrons
contribute separately to the electron-energy distribution, \ie that they can
be separated in the final state.

Our results for the lowest-order cross section 
$\rd\si/\rd \bar E'_{\Pe}$ and the asymmetry, 
together with the corresponding relative corrections, are shown
in \reftas{tab:nlccs} and \ref{tab:nlca}
as a function of the electron energy $\bar E'_{\Pe}$.
\btab
$$
\begin{array}{c@{\quad}c@{\quad}c@{\quad}c@{\quad}c@{\quad}c}
\hline\hline
\rule[-1.5ex]{0ex}{3.5ex}
{\bar E'_{\Pe} \ [\mathrm{GeV}]} &
{\dsidee
\left[\frac{\mathrm{mb}}{\mathrm{GeV}}\right]}&
\de_1 \ [\%]   &
\de_{\Pe} \ [\%]    & 
\de_{1\Pe}\ [\%]    \\
\hline  
   50.00  &       0.50884  &     0.64  &     0.42  &     1.06  \\
  100.00  &       0.25187  &     0.69  &     0.74  &     1.42  \\
  150.00  &       0.17774  &     0.58  &     0.66  &     1.24  \\
  200.00  &       0.14494  &     0.45  &     0.71  &     1.16  \\
  250.00  &       0.12796  &     0.33  &     0.95  &     1.27  \\
  300.00  &       0.11870  &     0.23  &     1.33  &     1.56  \\
  350.00  &       0.11378  &     0.18  &     1.41  &     1.59  \\
  400.00  &       0.11154  &     0.21  &     0.00  &     0.21  \\
  450.00  &       0.11107  &     0.43  &     0.00  &     0.43  \\
\hline\hline
\end{array}
$$
\caption{Lowest-order unpolarized cross section and relative
  percentage corrections for an NLC polarimeter 
  ($\bar E_\Pe=500\GeV$, $\bar E_\ga=2.33\eV$)}
\label{tab:nlccs}
\vspace*{2em}
$$
\newcommand{\m}{\phantom{-}}
\begin{array}{c@{\quad}c@{\quad}c@{\quad}c@{\quad}c@{\quad}c@{\quad}c}
\hline\hline
\rule[-1.5ex]{0ex}{3.5ex}
{\bar E'_{\Pe} \ [\mathrm{GeV}]} &
{A_{\mathrm{LR},0}} &
{\De A_{\mathrm{LR},1}} \times 100  &
{\De A_{\mathrm{LR},1e}} \times 100  &
\de_{A,1} \ [\%]    & 
\de_{A,1e} \ [\%]    \\
\hline  
   50.00  &    \m0.0094  &  -0.065  &      -0.201 &  -6.92 & -21.4 \\
  100.00  &     -0.5879  &  -0.003  &     \m0.288 & \m0.01 & -0.49 \\
  150.00  &     -0.7047  & \m0.113  &     \m0.417 &  -0.16 & -0.59 \\
  200.00  &     -0.6740  & \m0.195  &     \m0.470 &  -0.29 & -0.70 \\
  250.00  &     -0.5820  & \m0.240  &     \m0.522 &  -0.41 & -0.90 \\
  300.00  &     -0.4649  & \m0.261  &     \m0.530 &  -0.56 & -1.14 \\
  350.00  &     -0.3409  & \m0.271  &     \m0.397 &  -0.80 & -1.16 \\
  400.00  &     -0.2193  & \m0.284  &     \m0.284 &  -1.29 & -1.29 \\
  450.00  &     -0.1050  & \m0.309  &     \m0.309 &  -2.95 & -2.95 \\
\hline\hline
\end{array}
$$
\caption{Lowest-order asymmetry and corresponding absolute corrections 
 for an NLC polarimeter
  ($\bar E_\Pe=500\GeV$, $\bar E_\ga=2.33\eV$)}
\label{tab:nlca}
\etab
We separately give the relative corrections 
$\de_1$ to the cross sections of $\Pem\ga\to\Pem\ga(\ga)$, defined by
\beq
\frac{\rd\si_1}{\rd\bar E'_\Pe} = \frac{\rd\si_0}{\rd\bar E'_\Pe}(1+\de_1),
\eeq
and the corrections $\de_\Pe$ to the cross sections resulting from the
process $\Pem\ga\to\Pem\Pem\Pep$: 
\beq
\frac{\rd\si_{\Pe}}{\rd\bar E'_\Pe} = \frac{\rd\si_0}{\rd\bar E'_\Pe}\,\de_\Pe.
\eeq
Moreover, we include the sum of both relative corrections,
$\de_{1\Pe}=\de_1+\de_{\Pe}$.
In addition, we give the corrections to the asymmetry
with, $\De A_{\mathrm{LR,1e}}$, and without, $\De A_{\mathrm{LR,1}}$,
the contributions of the process $\Pem\ga\to\Pem\Pem\Pep$.
The corrections to the unpolarized cross section reach up to $1.7\%$
and contain an essential contribution from $\Pem\ga\to\Pem\Pem\Pep$. 
The absolute corrections to the asymmetry are 
about $5\times 10^{-3}$;
at least half of the effect comes from $\Pem\ga\to\Pem\Pem\Pep$. 
Since the relative corrections to the asymmetry amount to $1\%$ 
of the lowest-order asymmetry, 
they must be included in precision determinations of the beam polarization
at the NLC. For the total unpolarized cross sections we obtain
\beq
\si_0 =88.103\mba, \qquad \de_1 =  0.44\%, \qquad \de_{\Pe} = 0.29\%, 
\qquad \de_{1\Pe} =  0.72\%, 
\eeq
and the results for the related asymmetry are
\beq
\begin{array}[b]{rlrlrl}
A_{\mathrm{LR},0} &= -0.2593, \qquad & 
\De A_{\mathrm{LR,1}} &=0.00044, \qquad &
\De A_{\mathrm{LR,1\Pe}} &= 0.00047, \\
&& \de_{\mathrm{A,1}} &= -0.17\%, & \de_{\mathrm{A,1e}} &= -0.18\%.
\end{array}
\eeq

\bfi
\centerline{
\setlength{\unitlength}{1cm}
\begin{picture}(16,9)
\put(0,-1.5){\includegraphics{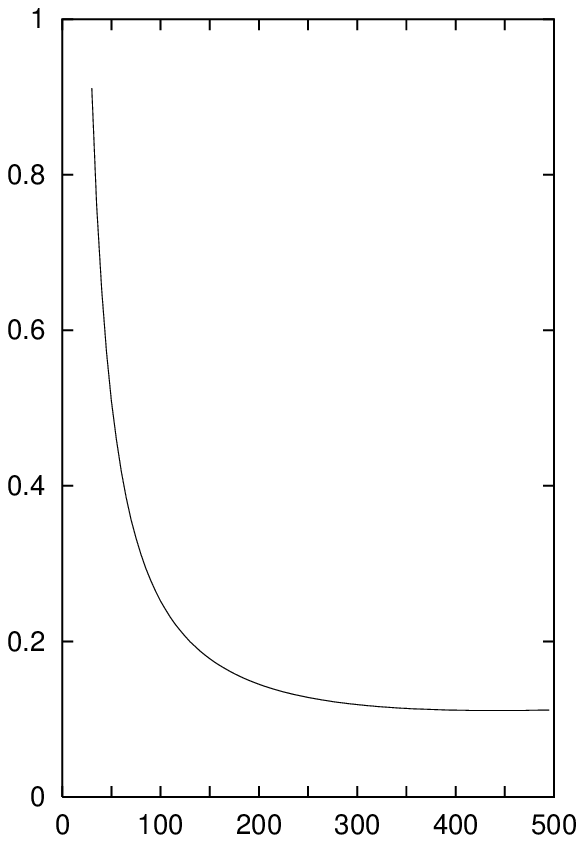}}
\put(8,-1.5){\includegraphics{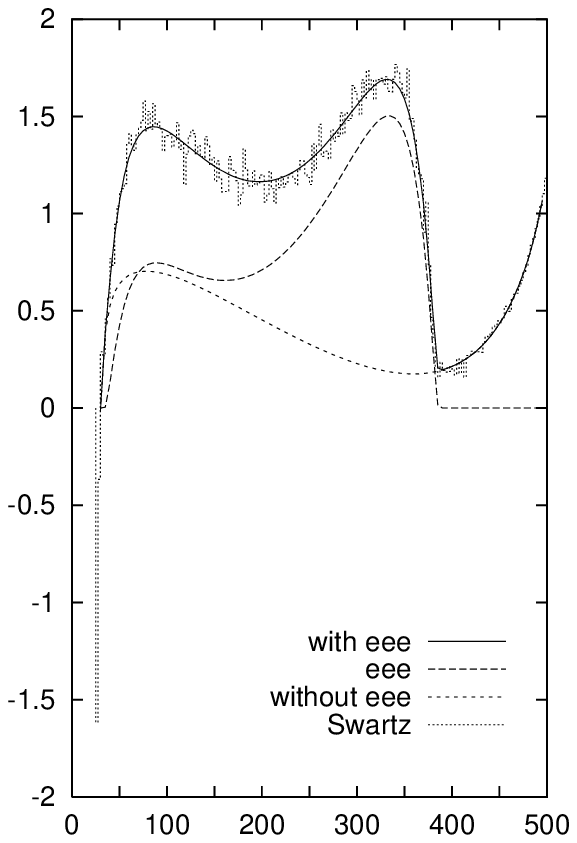}}
\put(0.5,5.5){\makebox(1,1)[c]{{$\disp\frac{\rd\si_0}{\rd\bar E'_{\Pe}} $}}}
\put(0,4.5){\makebox(2,1)[c]{{$ [\mathrm{mb}/\mathrm{GeV}]$}}}
\put(9.3,5.0){\makebox(1,1)[c]{$\de\  [\%]$}}
\put(5.0,-0.3){\makebox(1,1)[cc]{$\bar E'_\Pe/\mathrm{GeV}$}}
\put(13.0,-0.3){\makebox(1,1)[cc]{$\bar E'_\Pe/\mathrm{GeV}$}}
\end{picture}
}
\caption[]{Lowest-order unpolarized cross section (left) and
  corresponding percentage
  relative corrections (right) for an NLC polarimeter 
  ($\bar E_\Pe=500\GeV$, $\bar E_\ga=2.33\eV$)}
\label{fig:nlcsigma}
\efi
\bfi
\centerline{
\setlength{\unitlength}{1cm}
\begin{picture}(16,9)
\put(0,-1.5){\includegraphics{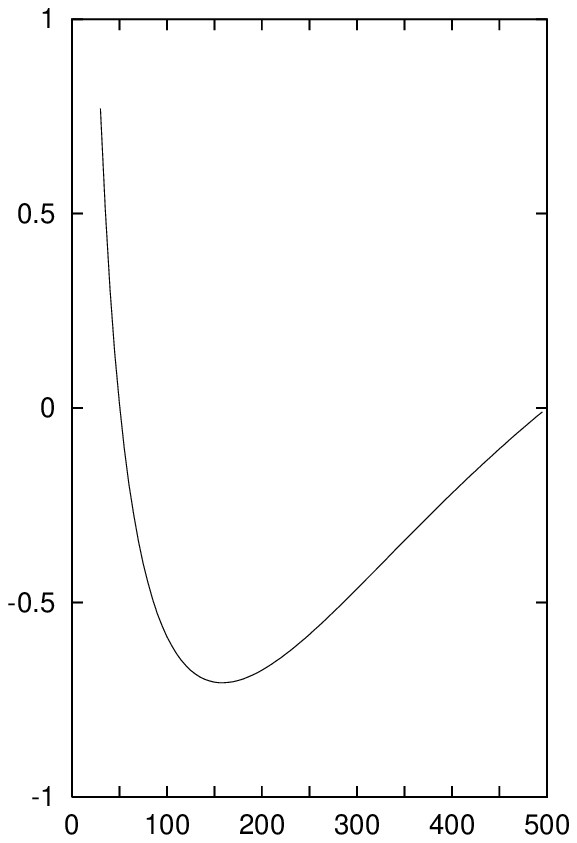}}
\put(8,-1.5){\includegraphics{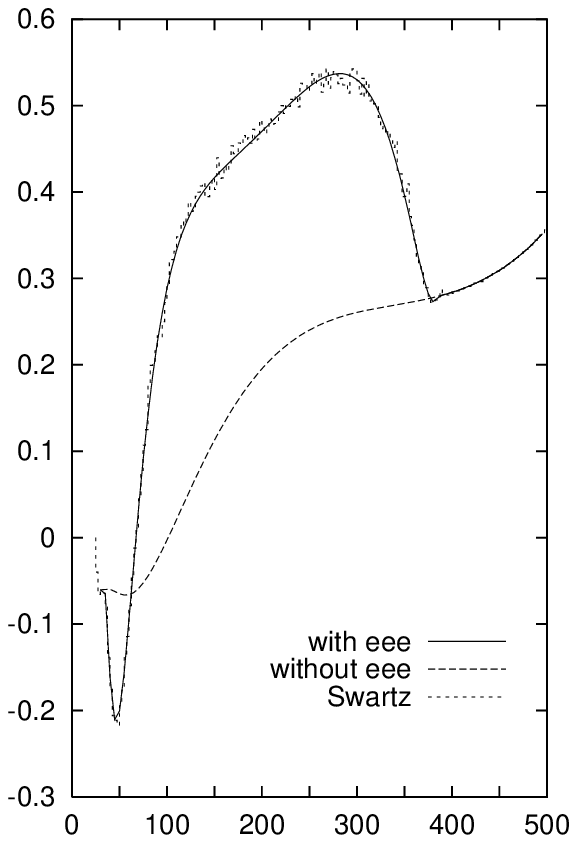}}
\put(1,5){\makebox(1,1)[c]{${A_{\mathrm{LR},0}}$}}
\put(9.0,5.3){\makebox(1,1)[c]{$100\times \De A_{\mathrm{LR}}$}}
\put(5,-0.3){\makebox(1,1)[cc]{$\bar E'_\Pe/\mathrm{GeV}$}}
\put(13,-0.3){\makebox(1,1)[cc]{$\bar E'_\Pe/\mathrm{GeV}$}}
\end{picture}
}
\caption[]{Lowest-order left--right asymmetry (left) and corresponding absolute
  corrections (right) for an NLC polarimeter 
  ($\bar E_\Pe=500\GeV$, $\bar E_\ga=2.33\eV$)}
\label{fig:nlcasymm}
\efi
Swartz \cite{sw97} 
has considered the same kind of NLC polarimeter. In
\reffis{fig:nlcsigma} and \ref{fig:nlcasymm} we compare our results
with his findings. We can fully confirm his results,
the variation of which is a measure of the 
Monte Carlo integration error. Note that we use a direct 
Monte Carlo integration of the differential cross
section, while he is using an event generator.

\section{Conclusion}
\label{se:concl}

In present and future \Pep\Pem~colliders the polarization of the
initial-state  particles plays a fundamental role. One way to measure this
polarization is by using Compton backscattering. Because of the use of
lasers, the relevant centre-of-mass (CM) energies are in the MeV region,
\ie of the order of the electron mass. 
The use of such a Compton polarimeter requires the knowledge of the
Compton process with sufficient accuracy,
\ie radiative corrections must be controlled.

The genuinely 
weak corrections are negligible, because the relevant energies are small 
with respect to the electroweak scale. We have calculated the
complete virtual, soft-photonic, and hard-photonic radiative
QED corrections to the process $\Pem\ga\to\Pem\ga$ for
arbitrarily polarized photons and electrons,
including the electron mass throughout. 
A very simple approximation for the corrections to the polarization
asymmetry for small CM energies has been presented.
We have also calculated the cross section for
$\Pem\ga\to\Pem\Pem\Pep$, which contributes to the corrections for
high-energy electron beams. For the processes
$\Pem\ga\to\Pem\ga\ga$ and $\Pem\ga\to\Pem\Pem\Pep$ we have
constructed analytical results for the matrix elements and determined
the integrated cross section by Monte Carlo integration.
We have presented all results in a form as simple as possible
and written all necessary formulas in a transparent way.
This facilitates the use of our results and computer programs
by our experimental colleagues.

We have applied our programs to three different situations: the CEBAF
polarimeter, which detects the energy of the scattered photon(s), the
SLD polarimeter, which detects the energy of the scattered electron,
and a polarimeter for a Next Linear Collider (NLC) that is similar to
the one for the SLD, which cannot distinguish between electrons from the 
processes $\Pem\ga\to\Pem\ga\ga$ and $\Pem\ga\to\Pem\Pem\Pep$. 
While the relative corrections to unpolarized cross sections are
suppressed for low CM energies, there is no such suppression
for the relative corrections to the polarization asymmetry.
{}For CM energies that are relevant 
to the CEBAF and SLD polarimeters
we find corrections to the asymmetry of the order
of a few $0.1\%$, whenever it is sizeable. For an NLC polarimeter with an
electron beam energy of $500\GeV$ we find corrections at the $1\%$
level; in this case the process $\Pem\ga\to\Pem\Pem\Pep$
contributes significantly.

We have compared our results for the SLD polarimeter
with those of H.~Veltman and Swartz, which
are not in mutual agreement. Between their results there are
significant deviations for the unpolarized cross section and small 
discrepancies for the polarization asymmetry.
We find complete agreement with the results of the generator COMRAD by 
Swartz for the NLC and SLD polarimeters. Thus, our code provides,
besides the generator COMRAD, a second completely independent program
for the electromagnetic corrections to polarized Compton scattering.
Our code can be adapted also to other situations, \eg to transversely
polarized $\Pepm$ beams. 

\appendix
 
\section*{Appendix}
\def\thesection{A}

\section*{Effects from a non-zero incident-beam angle}

In this paper we have consistently neglected the small angle $\alpha_c$
between the electron and photon beam axes. To justify this procedure, we
have calculated the lowest-order cross sections and the polarization 
asymmetry also for $\alpha_c$ different from zero. For angles $\alpha_c$
typically of order $10^{-2}$~rad, we find corrections to the results
for $\alpha_c=0$ that are at least one order of magnitude smaller than
the ones induced by the radiative corrections discussed in this paper.
In the following we briefly sketch the modifications that are necessary
to generalize the lowest-order formulas to non-zero $\alpha_c$.

There are two different origins of modifications induced by
$\alpha_c\ne 0$. 
{}First, of course the kinematics is changed. 
Secondly, owing to the acollinearity of the beam axes in the LAB frame,
the Lorentz transformation to the CM system is no longer a simple boost
along the electron direction of flight and thus involves 
also a spin rotation of the incoming electron.

We start by considering the kinematics. Identifying the $z$ axis 
in the LAB frame with the electron beam axis, the incoming momenta are
\beq
p^\mu = \bar E_\Pe(1,0,0,\bar\beta), \qquad
k^\mu =  \bar E_\gamma(1,0,-\sin\alpha_c,-\cos\alpha_c), 
\eeq
leading to 
\beq
s = \Me^2+2\bar E_\gamma \bar E_\Pe(1+\bar\beta\cos\alpha_c).
\label{eq:s_alpc}
\eeq
We decompose
the Lorentz transformation from the LAB frame to the CM system 
into three steps: 
step (i) is a rotation about the 
$x$ axis with angle $-\omega_1$,
which 
orients the spatial part of the total incoming momentum $p+k$ along 
the positive $z$ axis. Step (ii) is a boost along the (new) negative 
$z$ direction; its strength $\beb$ is chosen to lead us to the
CM system. Step (iii) is a rotation about the resulting 
$x$ axis with angle $\omega_2$, orienting the spatial parts of the 
incoming particle momenta
along the $z$ axis in the CM system. The components of $p$ and $k$ in
the CM system are the same as in \refeq{eq:momCMS}.
These considerations fix the parameters $\beb$ and 
$\omega_i$ to
\beq
\gab = \frac{1}{\sqrt{1-\beb^2}} = 
\frac{\bar E_\Pe+\bar E_\ga}{\sqrt{s}},
\qquad
\sin\omega_1 = 
\frac{\bar E_\ga\sin\alpha_c}{\beb(\bar E_\Pe+\bar E_\ga)},
\qquad
\sin\omega_2 = 
\frac{\bar\beta\sin\alpha_c}{\gab\beb(1+\bar\beta\cos\alpha_c)}.
\eeq
{}From this representation we can already read that $\omega_{1,2}$ are both of
${\cal O}(\alpha_c)$ and that they are additionally strongly suppressed by 
the small factors $\bar E_\ga/\bar E_\Pe$ and $\gab^{-1}$, respectively.

Since the rotational invariance about the electron beam axis is lost, we
have to include azimuthal angles $\bar\phi'_\Pe$ and $\phi'_\Pe$ in the
momentum $p'$ of the outgoing electron, as specified in 
\refeq{eq:eaeaa_mom}. The relation between the electron energy 
$\bar E'_\Pe$ in the LAB frame and the electron angles $\theta'_\Pe$,
$\phi'_\Pe$ in the CM system is given by
\beq
\bar E'_\Pe = E_\Pe\gab[1+\beb\beta
(\cos\omega_2\cos\theta'_\Pe-\sin\omega_2\sin\theta'_\Pe\sin\phi'_\Pe)].
\label{eq:eeo_ac}
\eeq
The CM quantities $E_\ga$, $E_\Pe$, and $\beta$ are related to $s$ as 
in \refeq{Eebeta} and \refeq{Mandelstam}.
In order to calculate unpolarized cross sections in the LAB frame, 
the spin-averaged squared amplitude, $\langle |{\cal M}|^2\rangle$, 
has to be known. Note that $\langle |{\cal M}|^2\rangle$ is a
Lorentz-invariant function of the Mandelstam variables $s$, $t$, $u$, and
thus that it does not depend on $\phi'_\Pe$ when expressed in terms of CM
quantities. Therefore, the unpolarized cross section $\sigma_0$ does not
explicitly depend on $\phi'_\Pe$ either.
{}For the distribution $\rd\si_0/\rd\bar E'_\Pe$ we obtain
\beq
\frac{\rd\sigma_0}{\rd\bar E'_\Pe} =
\frac{1}{E_\gamma \gab\beb} \int_0^{2\pi} \rd\psi \, 
\frac{\rd^2\sigma_0}{\rd\phi'_\Pe\rd\cos\theta'_\Pe}, 
\qquad
\bar E'_\Pe \gtrless E_\Pe\gab(1\mp\beb\beta),
\eeq
where the auxiliary variable $\psi$ determines the scattering angle
$\theta'_\Pe$,
\beq
\cos\theta'_\Pe = 
\left(\frac{\bar E'_\Pe-E_\Pe\gab}{E_\ga\gab\beb}\right)\cos\omega_2
+ \sqrt{1-\left(\frac{\bar E'_\Pe-E_\Pe\gab}{E_\ga\gab\beb}\right)^2}
\cos\psi\sin\omega_2,
\eeq
which plays the role of the scattering angle $\theta$ in \refeq{stu}.
The integration over $\psi$ is related to the integration over the
azimuthal angle $\phi'_\Pe$ in the CM system via
\beq
\sin\phi'_\Pe  =  \frac{1}{\sin\theta'_\Pe} \Biggl[
 \sqrt{1-\left(\frac{\bar E'_\Pe-E_\Pe\gab}{E_\ga\gab\beb}\right)^2}
\cos\psi\cos\omega_2
- \left(\frac{\bar E'_\Pe-E_\Pe\gab}{E_\ga\gab\beb}\right)\sin\omega_2
\Biggr].
\eeq
Owing to \refeq{eq:eeo_ac}
this integration includes an averaging process over $\theta'_\Pe$ when
$\bar E'_\Pe$ is kept fixed for $\alpha_c\ne 0$. For 
$\alpha_c=\omega_i=0$ the angle $\theta'_\Pe$ is fully determined by
$\bar E'_\Pe$ [see \refeq{eq:thboost}], and the integration over $\psi$
yields a factor $2\pi$. 

These considerations show that a finite $\alpha_c$
affects the unpolarized cross section 
through the CM energy $\sqrt{s}$, which is changed 
in ${\cal O}(\alpha_c^2)$ and enters
all kinematical variables, and via the 
auxiliary angles $\omega_i$, where the effects are of 
${\cal O}(\alpha_c)$ but additionally suppressed by at least a factor 
of $\gab^{-1}$.

{}For the calculation of polarized cross sections one has to take into
account the spin rotations of the electrons. In the following we only 
consider the polarization of the incoming electron and assume spin
summation for the final-state electron. Denoting Dirac spinors and 
amplitudes with spin $\sigma=\pm\frac{1}{2}$ 
in the CM system by $u(\alpha_c,\sigma)$
and $\M(\alpha_c,\sigma)$, respectively, the relations between
the amplitudes with $\alpha_c=0$ and $\alpha_c\ne 0$ read
\beq
\M(\alpha_c,\sigma) = \sum_{\tau=\pm} \,
 \M(\alpha_c=0,\tau)A_{\tau\sigma}, 
\qquad
A_{\tau\sigma} = \frac{1}{2\Me} 
\overline{u}(\alpha_c=0,\tau)u(\alpha_c,\sigma).
\eeq
The components of the unitary matrix $A$ are given by
\beq
A_{\pm\mp} =   
\frac{-\ri\Me\sin\frac{\alpha_c}{2}}
{\bar E_\Pe\sqrt{(1+\bar\beta)(1+\bar\beta\cos\alpha_c)}},
\qquad
A_{\pm\pm} = \sqrt{1-|A_{+-}|^2}.
\label{eq:Amat}
\eeq
The deviation of $A$ from the identity matrix is very small. 
The off-diagonal elements are of ${\cal O}(\alpha_c)$ times the
suppression factor of $\Me/\bar E_\Pe$, which is even smaller than
$\gab^{-1}$. The deviation of the diagonal elements from 1 is 
of the order of the square of the off-diagonal elements.

In conclusion, we find that the influence of an angle $\alpha_c\ne 0$ 
on the Compton cross sections and the polarization asymmetry
is at most of the relative order of $\alpha_c^2$ or $\alpha_c\gab$.
The above analytical considerations have also been checked numerically.

\section*{Acknowledgements}

We thank C.~Grosse-Knetter for participating in the early stage of this
work. Moreover,
we are grateful to C.~Cavata and M.L.~Swartz for helpful information
about the CEBAF and SLD polarimeters, respectively. 
We thank M.L.~Swartz also for providing us with explicit numbers 
of the results presented in \citere{sw97}.


\begin{thebibliography}{99}
\frenchspacing
\newcommand{\zp}[3]{{\sl Z. Phys.} {\bf #1} (19#2) #3}
\newcommand{\np}[3]{{\sl Nucl. Phys.} {\bf #1} (19#2) #3}
\newcommand{\phm}[3]{{\sl Phil. Mag.} {\bf #1} (19#2) #3}
\newcommand{\pl}[3]{{\sl Phys. Lett.} {\bf #1} (19#2) #3}
\newcommand{\pr}[3]{{\sl Phys. Rev.} {\bf #1} (19#2) #3}
\newcommand{\prl}[3]{{\sl Phys. Rev. Lett.} {\bf #1} (19#2) #3}
\newcommand{\prs}[3]{{\sl Proc. Roy. Soc.} {\bf #1} (19#2) #3}
\newcommand{\fp}[3]{{\sl Fortschr. Phys.} {\bf #1} (19#2) #3}
\newcommand{\cpc}[3]{{\sl Comput. Phys. Commun.} {\bf #1} (19#2) #3}
\newcommand{\ijmp}[3]{{\sl Int. J. Mod. Phys.} {\bf #1} (19#2) #3}
\newcommand{\nim}[3]{{\sl Nucl. Instr. Meth.} {\bf #1} (19#2) #3}
\newcommand{\nc}[3]{{\sl Nuovo Cimento} {\bf #1} (19#2) #3}
\newcommand{\vj}[4]{{\sl #1} {\bf #2} (19#3) #4}
\newcommand{\jcp}[3]{{\sl J. Comp. Phys.} {\bf #1} (19#2) #3}

\bibitem{sld} 
SLD Collaboration, K.\ Abe et al., \prl{70}{93}{2515}, 
\vj{}{73}{94}{25}, and \vj{}{78}{97}{2075}; \\
R.E. Frey, OREXP 97-03, hep-ex/9710016.

\bibitem{cebaf}
G. Bardin et al., {\it Conceptual Design Report of a Compton Polarimeter for
CEBAF Hall A}, 
SACLAY internal report DAPNIA-SPhN-96-14.

\bibitem{tjnaf}
J.P. Jorda, {\it A 4--8 GeV Compton Polarimeter for TJNAF}, SACLAY
internal report DAPNIA-SPhN-96-42.

\bibitem{tesla}
G. Bardin, C. Cavata and J.P. Jorda, {\it Compton Polarimeter Studies
for TESLA}, DESY-TESLA-97-03.

\bibitem{th50} W. Thirring, \phm{41}{50}{1193};\\
J.D. Bjorken and S.D. Drell, {\it Relativistic Quantum Fields}
(McGraw-Hill, New York, 1965).

\bibitem{di97} S. Dittmaier, \pl{B409}{97}{509}.

\bibitem{br52} L.M.\ Brown and  R.P.\ Feynman, \pr{85}{52}{231}.

\bibitem{ma52} F.\ Mandl and T.H.R.\ Skyrme, \prs{A215}{52}{497}.

\bibitem{mi72} K.A.~Milton, W.~Tsai and L.L.~De~Raad,
\pr{D6}{72}{1411} and {\bf D6} (1972) 1428.

\bibitem{go89} A.~G\'ongora-T. and R.G.~Stuart, \zp{C42}{89}{617}.

\bibitem{ve89} H.~Veltman, \pr{D40}{89}{2810}; 
E:~\vj{ibid.}{D42}{90}{1856}.

\bibitem{sw97} M.L.~Swartz, SLAC-PUB-7701, hep-ph/9711447.

\bibitem{de93} A.~Denner and S.~Dittmaier, \np{B407}{93}{43}.

\bibitem{di94} S.~Dittmaier, \np{B423}{94}{384}.

\bibitem{di98} S. Dittmaier, CERN-TH/98-143, hep-ph/9805445.

\bibitem{ADHab} A. Denner, \fp{41}{93}{307}.

\bibitem{pa79}  G.\ Passarino and M.\ Veltman, \np{B160}{79}{151}.

\bibitem{th79} G.\ 't Hooft and M.\ Veltman, \np{B153}{79}{365}; \\
W.\ Beenakker and A.\ Denner, \np{B338}{90}{349}.

\bibitem{FA} J.~K\"ublbeck, M.~B\"ohm and A.~Denner, \cpc{60} {90}{165}; \\
H.~Eck and J.~K\"ublbeck, {\it Guide to FeynArts 1.0\/}, 
University of W\"urzburg, 1992.

\bibitem{FC} R.~Mertig, M.~B\"ohm and A.~Denner, \cpc{\bf 64}{91}{345}; \\
R.~Mertig, {\it Guide to FeynCalc 1.0\/}, University of W\"urzburg, 1992.

\bibitem{de95} A.~Denner, S.~Dittmaier and M.~Strobel, \pr{D53}{96}{44}.

\bibitem{le78} G.P.~Lepage, \jcp{27}{78}{192};
Cornell University preprint, CLNS-80/447.

\bibitem{kl85} 
{}F.A. Berends, P.H. Daverveldt and R. Kleiss, \np{B253}{85}{441}; \\
R. Kleiss and W.J. Stirling, \np{B262}{85}{235} and \pl{179B}{86}{159}; \\
R. Kleiss, \zp{C33}{87}{433}.

\bibitem{LEPEWWG97}
The LEP collaborations ALEPH, DELPHI, L3, OPAL, the LEP Electroweak
Working Group and the SLD Heavy Flavor Group (D. Abbaneo et al.),
CERN-PPE/97-154.

\end{thebibliography}
\end{document}